\documentclass{aa}

\usepackage{txfonts}
\usepackage{graphicx}
\usepackage{lscape}
\usepackage{longtable}
\usepackage{color}

\begin{document}

\title{Variability of massive stars with known spectral types \\ in the Small Magellanic Cloud using 8 years of OGLE-III data}
\author{M. Kourniotis\inst{1,2} \and A.Z. Bonanos\inst{1} \and I. Soszy\'{n}ski\inst{3} \and R. Poleski\inst{3,4} \and G. Krikelis\inst{2} \and A. Udalski\inst{3} \and M.K. Szyma\'{n}ski\inst{3} \and M. Kubiak\inst{3} \and G. Pietrzy\'{n}ski\inst{3,5} \and {\L}. Wyrzykowski\inst{3} \and K. Ulaczyk\inst{3} \and S. Koz{\l}owski\inst{3} \and P. Pietrukowicz\inst{3}}

\institute{IAASARS, National Observatory of Athens, GR-15236 Penteli, Greece \\ \email{mkourniotis@astro.noa.gr,bonanos@astro.noa.gr} \and Section of Astrophysics, Astronomy and Mechanics, Faculty of Physics, University of Athens, Panepistimiopolis, GR15784 Zografos, Athens, Greece \and Warsaw University Observatory, Al. Ujazdowskie 4, 00-478 Warszawa, Poland \and Department of Astronomy, Ohio State University, 140 W. 18th Ave., Columbus, OH 43210, USA \and Universidad de Concepci\'{o}n, Departamento de Astronomia, Casilla 160-C, Concepci\'{o}n, Chile} 

\date{}
\authorrunning{Kourniotis et al.}
\titlerunning{Variability of massive stars in the SMC}

\abstract{We present a variability study of 4646 massive stars in the Small Magellanic Cloud (SMC) with known spectral types from the catalog of \cite{Bonanos10} using the light curves from the OGLE-III database. The goal is to exploit the time domain information available through OGLE-III to gain insight into the processes that govern the evolution of massive stars. This variability survey of massive stars with known spectral types is larger than any previous survey by a factor of 7. We find that 60\% of our sample (2766 stars) show no significant variability and 40\% (1880 stars) exhibit variability distributed as follows: 807 stars display low-amplitude stochastic variability with fluctuations in $I-$band of up to 0.05 mag, 443 stars present irregular variability of higher amplitude (76\% of these are reported as variables for the first time), 205 are eclipsing binaries (including 101 newly discovered systems), 50 are candidate rotating variables, 126 are classical Cepheids, 188 stars exhibit short-term sinusoidal periodicity ($P < 3$ days) making them candidate ``slowly pulsating B stars'' and non-radial Be pulsators, and 61 periodic stars exhibit longer periods. We demonstrate the wealth of information provided in the time domain, by doubling the number of known massive eclipsing binary systems and identifying 189 new candidate early-type Be and 20 Oe stars in the SMC. In addition, we find that $\sim80\%$ of Be stars are photometrically variable in the OGLE-III time domain and provide evidence that short-term pulsating stars with additional photometric variability are rotating close to their break-up velocity.}

\keywords{binaries: eclipsing -- galaxies: individual (SMC) -- stars: early-type -- stars: variables: general -- stars: massive -- stars: emission-line, Be}

\maketitle 

\section{Introduction}
\label{section:intro}
The intrinsic variability of massive stars ($\gtrsim$ 8~M$_{\sun}$) is prominent from their main-sequence lifetime to their end-state evolution. Very massive stars exhibit strong photospheric winds as a result of radiation pressure \citep{Kudr00}, yielding high mass-loss rates, which can cause line profile variability and possibly brightness variations \citep[e.g.,][]{Rich11}. Given the dependence of line driven winds and consequently mass loss rates on metallicity, a significant fraction of early-type stars close to the main sequence exhibit fast rotation in low-metallicity environments. These so-called ``Be stars'' eject mass that accumulates in a circumstellar disk, the establishment and dissipation of which results in spectroscopic and photometric variability. Other classes of early B-type stars, such as $\beta$ Cephei variables and slowly pulsating B stars (SPBs), display short-term periodic modulation in their light curves as a result of radial pulsations driven by the $\kappa$-mechanism, which is due to the iron-group opacity bump. Low metal abundance appears to narrow or even make the instability strip of B-type pulsators vanish \citep{Miglio07}. In particular, after adopting a chemical mixture representative of B stars in the SMC, no $\beta$ Cephei stars are expected at $Z<0.007,$ while no SPBs are expected at $Z<0.005$ \citep{Salmon12}.

On the other hand, the extrinsic variability of massive stars is mostly prominent in the case of eclipsing systems. The high binarity fraction of early-type OB stars \citep{Sana12,Chini12} makes the detection of photometric and spectroscopic variability common. Eclipsing binaries provide a geometrical method for deriving accurate fundamental parameters of stars \citep{Andersen91,Torres10}. Photometry combined with spectroscopy yields the physical radii of the systems, individual masses, and therefore, the surface gravities. Furthermore, effective temperatures and luminosities can be estimated, enabling an accurate and direct determination of the reddening and the distance to OB-type stars in galaxies of the Local Group \citep[e.g.,][]{Bonanos06,Bonanos11}. Measuring the fundamental parameters of massive stars accurately is of prime importance for constraining models of the formation and evolution of massive stars \citep[e.g.,][]{Bonanos09}. Eclipsing binary measurements also provide the most robust values for the upper mass limit \citep[e.g.,][]{Bonanos04,Rauw04}.

The main goals of this work are to identify new eclipsing binaries with massive components and to explore the Be phenomenon via photometric variability. Classical Be stars are B-type stars with luminosity classes III-V, which rotate close to their critical limit \citep{Porter03,Rivi13}. They exhibit hydrogen Balmer emission, as well as emission of He I, Fe II, and sometimes Si II and Mg II, with double-peaked profiles, indicating the presence of a circumstellar, gaseous, quasi-Keplerian disk. Spectroscopy confirms the Be nature of a star; however, the transience of the phenomenon means that the method is not definitive. The modification of the disk owing to the fast rotation possibly combined with non-radial pulsations \citep{Riv98} is widely accepted as producing outbursts and fluctuations in brightness. Several explanations of the formation of the disk have been proposed, such as\ \cite{Bjork93}, \cite{Cass02}, and \cite{Riv01}. It has also been established that metallicity plays a major role in the frequency and the intensity of the Be phenomenon. Low metal abundance results in low mass-loss rates since the radiation pressure that is responsible for driving the stellar winds acts on the metal lines \citep{Vink01}, therefore angular momentum loss remains low, and the star reaches a high $\Omega/\Omega_{c}$ ratio, triggering the Be phenomenon. As a result, the fraction of early-Be stars in the SMC is found to exceed the fraction in the Galaxy and the Large Magellanic Cloud (LMC) \citep{McSwain05,Wisniewski06,Martayan10}, specifically in the second half of their main-sequence lifetimes. In addition, the duration and amplitude of the outbursts as ejections of matter from Be stars, appear longer and larger in the SMC than in the LMC \citep{Sabogal05}.

Over the past two decades, the microlensing experiments MACHO \citep{Alcock93}, the Optical Gravitational Lensing Experiment II \citep[OGLE-II;][]{Udalski97}, and EROS \citep{Aubourg95} have provided a database of light curves of a large number of massive stars located in the Magellanic Clouds and the Galaxy, motivating detailed studies of Be stars \citep{Keller02,Menn02,deWit03,Sabogal05,Sabogal08} and other types of massive stars. Furthermore, surveys with the All Sky Automated Survey \citep[ASAS;][]{Pojmanski02} and the Hipparcos mission have motivated a couple of studies on variability as a function of spectral type. \cite{Szczygiel10} exploited ten years of photometry provided by ASAS to probe for variability in $\sim$600 massive stars in the LMC. They focused on bright, late spectral-type stars and found 117 variables of which $\sim$$73\%$ were red supergiants. \cite{Hubert98} used the four-year Hipparcos photometry of 289 Galactic Be stars with spectral types from O9.5 to A0, consisting of 110 epochs, to detect short-term, low-amplitude, periodic variability in $\sim$86\% of early-Be stars. They also excluded the presence of short-lived outbursts in late-Be type galactic stars. 

The goal of this paper is to take advantage of the eight-year, wide-field monitoring of the OGLE-III project, exploit the time domain and correlate variability with the spectral type of massive stars in the SMC. We used the catalog of 5324 massive stars with known spectral types compiled by \citet{Bonanos10} from the literature. The paper is organized as follows. In Sect. 2, we give a brief description of the input spectral catalog and the OGLE-III survey that provided the light curve database. In addition, we describe the matching procedure and the constraints that resulted in the final list of stars studied. In Sect. 3, we present the method used to separate the ``real'' from the ``spurious'' variables and to search for periodic signals. In Sect. 4, we report our results on periodic and irregular variables. In Sect. 5, we discuss our results and derive statistics for stars exhibiting clear photometric variability. In Sect. 6, we summarize the main results of the present work.

\section{Input catalogs}
We used the catalog of 5324 massive stars in the SMC compiled by \cite{Bonanos10}, which contains their name, coordinates (J2000.0), spectral classification (typically accurate to one spectral sub-type and one luminosity class), and reference. The catalog consists of 12 Wolf Rayet, 277 O-type, 3069 early B-type, 602 late B-type, 1222 AFG-type, 107 K and M-type stars, 34 stars classified as ``Be'' (including 21 X-ray binaries) and treated in the present work as early B-type stars, and 1 supergiant B[e] star.

Our goal is to match the stars from the catalog of \cite{Bonanos10} to stars with light curves from the OGLE-III database. The OGLE-III survey covered 14 square degrees  \citep[41 fields;][]{Udalski08}, which are centered on the densest regions of the SMC. The observational data were obtained between 2001 and 2009 with the 1.3 m Warsaw telescope at Las Campanas Observatory in Chile. The telescope was equipped with an eight-chip CCD mosaic camera with 8192$\times$8192 pixels and a field of view of about 35$\times$35 arcmin \citep[details on the instrumentation setup can be found in][]{Udalski03}. About 90\% of the observations were collected in the Cousins $I-$band, while 10\% were obtained in the Johnson $V$ photometric band.

\subsection*{Matching procedure}
For each of the 5324 stars in the catalog, we examined all the OGLE-III light curves matched within a radius of $2\arcsec$, given the $1\arcsec$ accuracy in the coordinates of the stars of the input catalog. A total of 582 stars had no match since they correspond to stars not covered by the OGLE-III fields or to very bright stars that are saturated in the OGLE-III images, including all 108 K and M supergiants of the catalog. We found that 2188 stars only have one match. For 2554 stars, more than one light curve was detected within $2\arcsec$. In these cases, we selected the brightest source as the correct match unless there was a source closer than $1\arcsec$ that was within 0.5 mag of the brightest match, so as to avoid selecting faint instrumental artifacts that usually lie in the vicinity of bright stars. 

In total, 485 close matches were selected using this criterion. To confirm our selection, we compared the mean $I$ and $V$ values of the candidate light curves to the respective ones from the Magellanic Clouds Photometric Survey \citep[MCPS;][]{Zaritsky02}, as provided in \cite{Bonanos10} when available. We set a differential limit of 0.5 mag, below which the selected light curve was considered to be the correct match, in order not to reject counterparts of variable stars with outbursts or deep eclipses. A total of 70 stars were found to exceed the particular limit in at least one band. We kept five that proved to be eclipsing binaries, 4 Cepheids and 14 stars with significant irregular variability, as we discuss later. The remaining 47 were excluded from our study, since they presented constant light curves or even low-amplitude variability across the OGLE-III monitoring period, which could not justify the discrepancy of 0.5 mag or more. However, we caution that the photometry of MCPS was conducted before the OGLE-III survey, and therefore, some of these rejected stars might have truly exhibited variable behavior in the past.   

\begin{figure*} 
\centering
\includegraphics[width=5.5in]{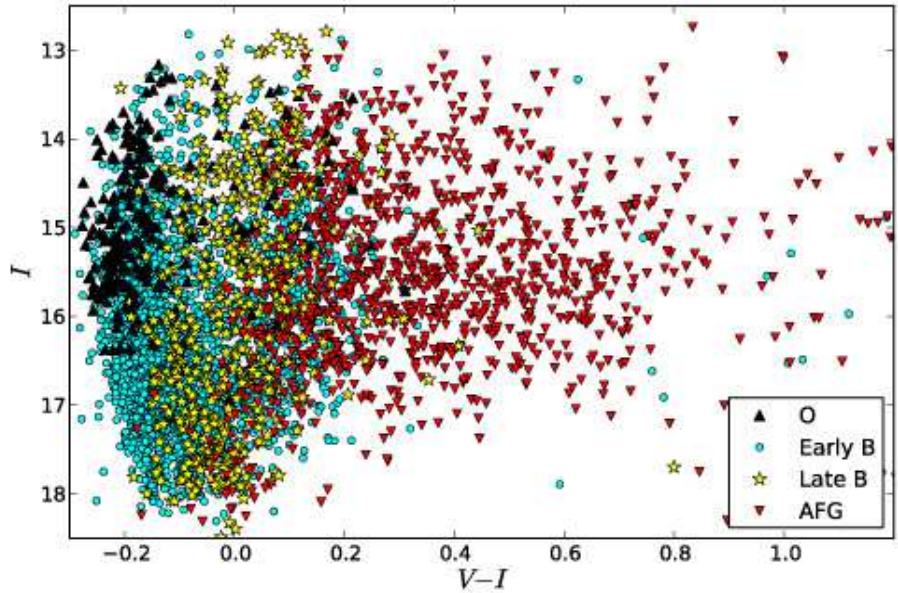}
\caption{$I$ vs. $V-I$ color-magnitude diagram for the 4614 stars from our sample of 4646 stars that have both $I$ and $V-$band OGLE-III photometry. The spectral types are labeled as follows: O stars as black triangles, early-B stars as cyan circles, late-B stars as blue stars, and AFG-type stars as red inverted triangles. Only 1.3\% of early-B type stars exhibit very red colors due to contamination.}
\label{fig01}
\end{figure*}

We ended up with a catalog of 4646 stars with $I-$band magnitudes between $12.6-19.4$, having also excluded 15 lightcurves with fewer than 100 epochs, because the median number of epochs for the matched light curves in the $I-$band is 698, as well as 31 lightcurves that correspond to bright stars close to saturation or to known types of artifacts in the OGLE photometry. We also excluded three stars verified as double entries\footnote{NGC346-016/2dFS5100, NGC330-110/SMC5\_037341 and OGLE-td9175323/Hodge53h53-47} in the input catalog. The color-magnitude diagram of the catalog is presented in Fig.~\ref{fig01}. It consists of 4614 stars, since 32 stars were not measured in the $V-$band. We note that the very reddened colors ($V-I$ \textgreater ~0.5 mag) of some early and late B-type stars are due to contamination by hot spots in the interstellar medium, as recently shown by \citet{Sheets13} and \citet{Adams13}. The percentage of early B-type stars affected by contamination is only 1.3\%.

\section{Analysis}
\begin{figure*}
\centering
\includegraphics[width=5.5in]{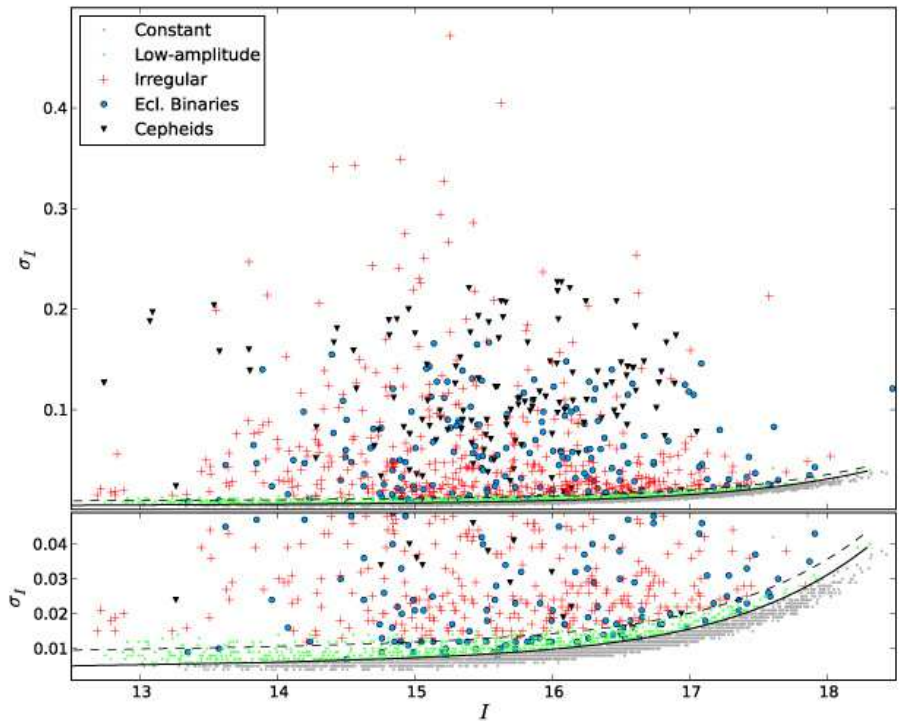}
\caption{$\sigma_{I}$ vs. $I$ diagram for the 4646 stars studied in this work. The solid curve is used to separate real from spurious variables \citep[as proposed by][]{Graczyk10}, while the region between the solid and the dashed curves defines the low-amplitude variables. A zoomed region is presented in the lower panel. The irregular variables (crosses), the Cepheids (inverted triangles), and most of the EBs (open circles) lie above the dashed curve.}
\label{fig02}
\end{figure*}

Our analysis of the 4646 OGLE-III light curves is based on the four statistical moments of the $I-$band light curve: the mean magnitude, standard deviation, skewness, and the kurtosis. We used a 2$\sigma$ cut over the mean to remove possible outliers from each light curve and recalculated the statistics. To distinguish between real variables and non-variables, we used the $\sigma_{I}$ vs.\ $I$ plane where $\sigma_{I}$ is the standard deviation of the $I-$band light curve points (Fig.~\ref{fig02}). In this plane, the real variables are defined according to~\cite{Graczyk10} for the OGLE-II database of light curves in the LMC as those that lie above their proposed, empirical curve. Our goal is to search for significant variability so we decided to add a shifted, elevated curve and classified 935 stars that lie above as high-amplitude variables, 945 stars that were found between the two curves as low-amplitude variables, and 2766 stars that lie below the primary curve as constant. In fact, since the primary curve is empirical, we expect partial overlap between the low-amplitude region and that of the non-variables. Our low-amplitude variables present variability $\Delta I$ $\lesssim$ 0.05 mag.

\begin{figure*}
\centering
\includegraphics[width=5.5in]{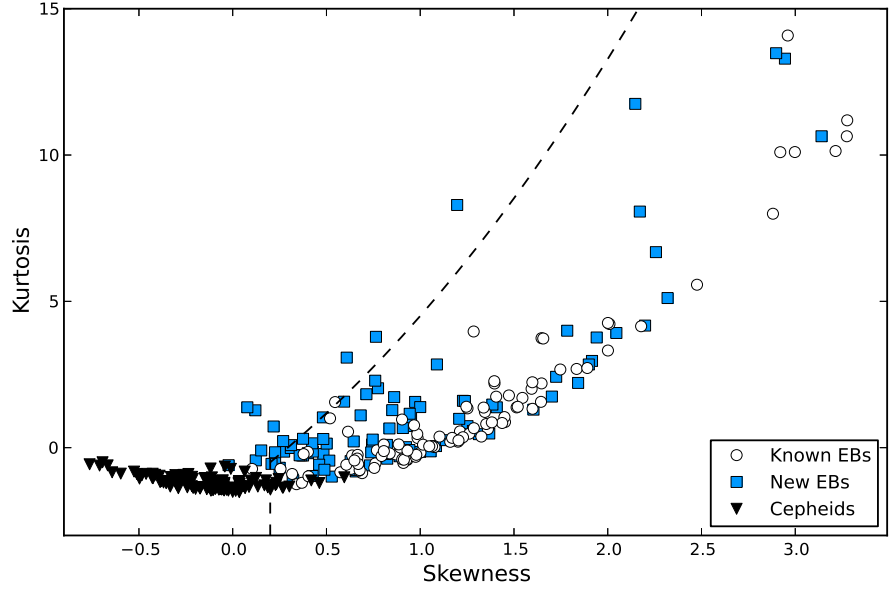}
\caption{Skewness-Kurtosis diagram for the eclipsing binaries and the Cepheids identified in the spectral catalog. New eclipsing binaries are indicated with blue squares, and known eclipsing binaries with circles. Eclipsing binaries are expected to be found below the dashed curve for skewness values higher than 0.2, as defined by \cite{Graczyk10}. However, the method was found to miss systems with sparsely sampled eclipses (above the dashed curve) or with rather symmetric light curves (close to (0,0)).}
\label{fig03}
\end{figure*}

For the detection of eclipsing binaries (EBs), we followed the method introduced by~\cite{Graczyk10} based on the skewness and the kurtosis of the distribution of magnitudes. We initially focused on those stars in the region of the skewness-kurtosis diagram where we expect eclipsing binaries to lie. Cepheids tend to be more symmetric with negative kurtosis, and as a result they are found in the lower region of this diagram. Figure~\ref{fig03} shows the location of the Cepheids from our catalog and both our newly discovered and the known eclipsing binaries on the skewness-kurtosis diagram. The light curves were searched for periodicity using the Analysis of Variance (AoV) algorithm \citep{Schwarz89} in the frequency range of 0.0001-7 cycles per day and were folded to the period corresponding to the frequency with the highest peak in the power spectrum. The discovery of numerous short- and long-term sinusoidally varying periodic stars located close to the $(0,0)$ point of the skewness-kurtosis diagram motivated us to extend our periodic search. Eventually, all low-amplitude variables, as well as some of the high-amplitude variables, which were strongly suspected of being periodic after visual inspection of their raw light curves, were inspected with the period search algorithm. This yielded a significant number of short- and long-period variables representative of pulsating and rotating variables, as discussed in Sect. \ref{sec:res}. In addition, we identified 13 eclipsing binaries that were found to lie outside the borders of their expected region since they were either very detached with few downward outliers or had rather symmetric light curves. 

\section{Results}
\label{sec:res}

\subsection {Periodic variables}

\subsubsection {Eclipsing binaries}
\label{sec:ecl}
The discovery of new eclipsing binaries with high-mass components is the first step toward populating the high-mass end of the mass-radius relation with accurate measurements derived dynamically using double-lined spectroscopic eclipsing binaries. During the periodic search described in the previous section, we identified 205 eclipsing binaries. A total of 20 of them present a displacement of the secondary eclipse with time, which is an indication of apsidal motion. In addition, the EB 2dFS2780 ($P=5.602226$ days) was found to exhibit additional long-term periodicity ($P2=148.80952$ days) and is a candidate double-periodic variable. This class of variables was first introduced in the SMC by \cite{Mennickent03} and has also been investigated in the LMC \citep{Mennickent03, Poleski10}. Their enigmatic nature reveals a short-term period due to the orbital motion of a semi-detached binary and a long-term period that lasts $\sim33$ times the orbital period. The long-term period probably stems from the modulation of a circumbinary disk \citep{Mennickent08}. We also identified two transient EBs 2dFS2090 and 2dFS3560 caused by orbital precession. The former is an EB with a period of 1.420025 days that only presents an additional EB observed in the middle of the OGLE-III time domain. The latter is a transient EB with a period of 6.223004 days, which appears to vanish close to the end of the monitoring period.

\begin{figure*}
\centering
\includegraphics[width=7in]{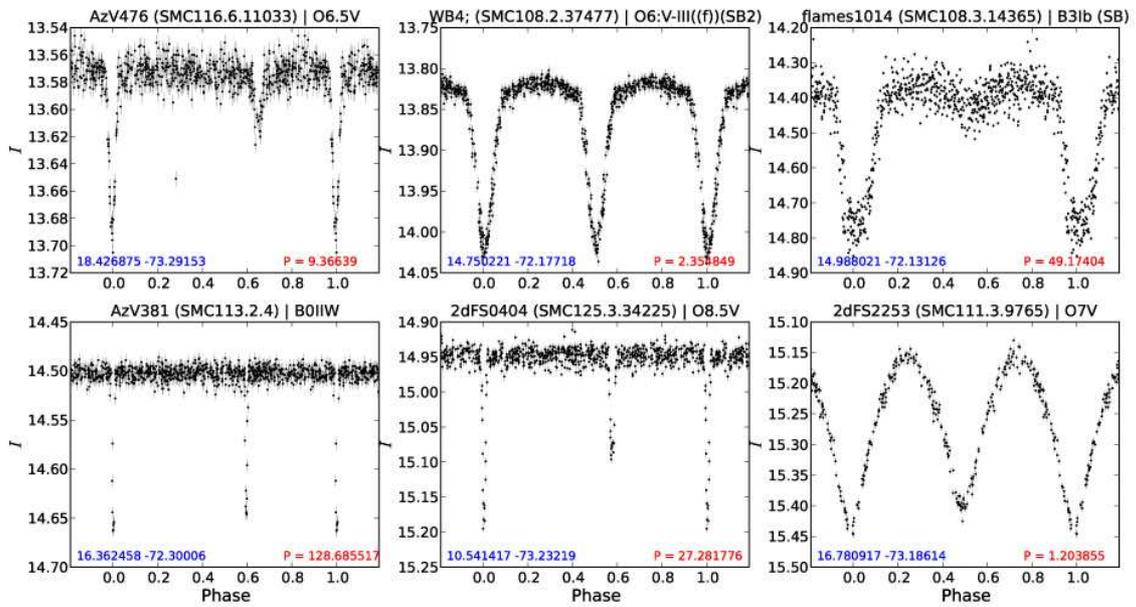}
\caption{Phased light curves of 6 newly discovered eclipsing binaries, along with their names, OGLE-III field and database numbers, spectral types, coordinates (J2000.0), and derived periods (in days), demonstrating the range of spectral types, luminosity classes, and binary configurations included in our sample. Error bars are indicated on each point.}
\label{fig04}
\end{figure*}

We searched each EB individually in the VizieR catalog, using a $3\arcsec$ search radius, and found 101 systems without any previous reference to their eclipsing nature. Of these 101 systems, 8 are O-type binaries, 85 early B-type,  and 7 late B-type, and only one has an F-type. In addition, nine of the new EBs have already been classified as spectroscopic binaries by \cite{Evans06} and Hunter et al. (in prep.). Table~\ref{tab01} presents the main characteristics of the 101 new EBs sorted by spectral type, listing the star name(s), spectral type, coordinates in degrees (J2000.0), values of the skewness and kurtosis, the OGLE-III database number, mean magnitudes in the $I$ and $V$ bands, and period (days). We flag whether it is a reported single, double, or undefined spectroscopic binary, and then note the systems with apsidal motion, the transient EBs, and the double-periodic variable. Representative phased light curves of six newly discovered eclipsing binaries are presented in Fig.~\ref{fig04}, along with their names, coordinates, spectral types, and derived periods. Phase zero corresponds to the deeper eclipse.

\begin{figure*}
\centering
\includegraphics[width=7in]{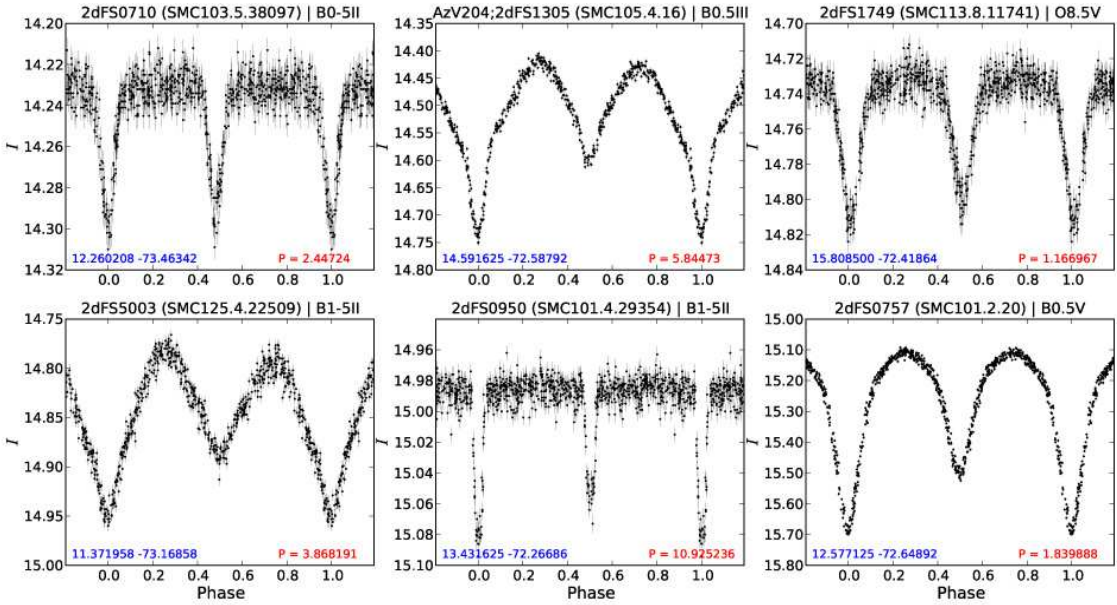}
\caption{Phased light curves of 6 known eclipsing binaries, as in Fig.~\ref{fig04}. For many of these systems, their spectral types are presented for the first time.}
\label{fig05}
\end{figure*}

The OGLE-III observations provide us with high-quality, long-term light curves for the 104 known EBs, so that our derived periods in many cases are more precise than those reported in the literature. Figure~\ref{fig05} presents OGLE-III light curves of six previously known eclipsing binaries. Table~\ref{tab02} presents the 104 known eclipsing binaries as in Table~\ref{tab01}, along with the most recent reference to the work that reports the eclipsing nature. The input catalog contains 45 EBs that have already been analyzed by \citet{Hild05} and \citet{Harries03}. We have also identified 59 EBs, using the VizieR catalog, which have been reported in previous works, mainly by \citet{Wyrz04} and \citet{Faccioli07}. Six of these have been spectroscopically confirmed. We also note the case of star 2dFS1636, which was reported to have composite features of A and F type stars \citep{Evans04}. We found it to be an eclipsing binary, which suggests binarity as a natural explanation for stars exhibiting such composite features in their spectra. This star was reported by \citet{Faccioli07} as an eclipsing binary, however we are the first to make the connection to its spectral type.  

\subsubsection {Candidate rotating variables / Overcontact binaries}
\label{sec:rot}

\begin{figure*}
\centering
\includegraphics[width=6.5in]{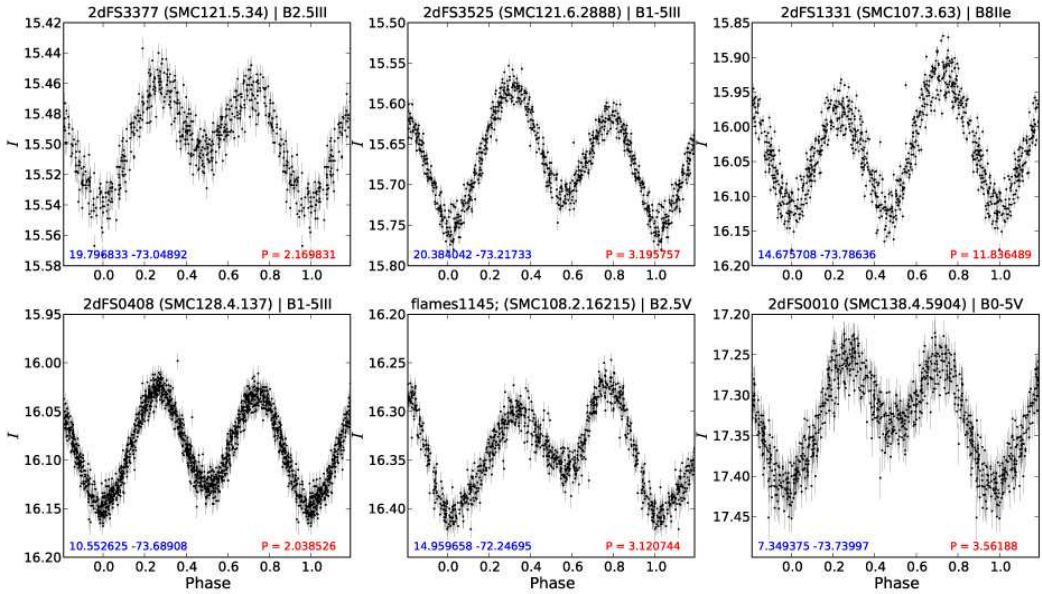}
\caption{Phased light curves of 6 candidate rotating variables, as in Fig.~\ref{fig04}. Sinusoidally periodic variables with unequal minima are included in this category.}
\label{fig06}
\end{figure*}

Rotating variables owe their variations in brightness to non-uniform surface features or to the non-spherical shape of distorted components
within binary systems. The most common classes of rotating variables include the chemically peculiar stars (CPs) and ellipsoidal variables. CPs are stars in the upper part of the main-sequence with B or A spectral types, which present an inhomogeneous surface distribution of chemical
elements and yield near-sinusoidal light curves as a result of stellar rotation. Variability is often attributed to strong
and stable magnetic fields \citep[e.g.,][]{Luftinger10,Krticka12}, although cases with weak magnetic fields have been reported as well \citep[e.g.,][]{Krticka09,Shulyak10}. The amplitude of the light variations decreases with increasing wavelength \citep{Catalano91}. Ellipsoidal variables are close binaries with non-spherical components due to tidal interactions, whose folded light curves present sinusoidal morphology \citep{Soszynski04}. In some cases, overcontact eclipsing binaries can produce ellipsoidal-like light curves \citep[although of higher amplitude;][]{Prsa11} especially if their light curves are noisy. Since these categories of variables produce identical light curves that can also be misclassified as pulsating stars, we decided to flag a star as a ``candidate rotating variable'' if it presents a near-sinusoidal light curve with unequal minima/maxima, since this is an indication of tidal distortion effects for an ellipsoidal star or uneven surface distribution of elements. We also include stars with equal minima that have been spectroscopically identified as binaries. A periodic variable that does not meet the requirements of an eclipsing binary or a candidate rotating variable has been flagged as a short or a long periodic variable, as discussed in Sect. \ref{sec:other}. In Table~\ref{tab03} we list the parameters of the 50 candidate rotating variables or overcontact binaries, and we report the cases where the star has been classified in previous works as an eclipsing binary. Figure~\ref{fig06} presents the OGLE-III phased light curves of six candidate rotating variables.

\subsubsection {Cepheids}
\label{sec:ceph}
\begin{figure*}
\centering
\includegraphics[width=5.5in]{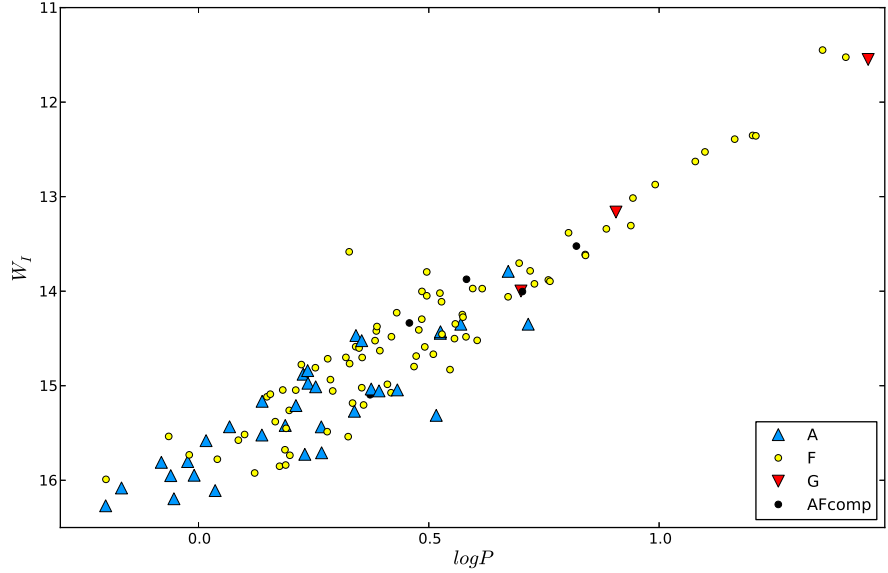}
\caption{Period-luminosity diagram for the Cepheids in the SMC that are included among our variables. Green triangles represent A-type stars, yellow circles F-type, red inverted triangles G-type, and black circles stars with composite features of A and F-type (AFcomp), which we propose to be binary systems.}
\label{fig07}
\end{figure*}

The Cepheids in the SMC have already been the subject of extended study by \cite{Sosz10} using photometry provided by OGLE-III. However, the spectroscopic study of Cepheids has been more limited, even though it is crucial for understanding the properties of the stars themselves, their use as distance indicators \citep[e.g.,][]{Romaniello08}, and as tracers of abundance gradients in galaxies \citep[e.g.,][]{Andrievsky02}. We identified 126 classical Cepheids among the stars in our catalog. They have already been reported in \cite{Sosz10} and we provide spectral types for these, finding 33 A-type stars, 85 F-type stars, 3 G-type stars, and 5 stars with composite features of both A and F stars \citep{Evans04}, which are likely non-eclipsing binaries. Table~\ref{tab04} lists the main characteristics of the Cepheids, as in Table~\ref{tab01}. Figure~\ref{fig07} presents the period-luminosity (PL) diagram for the matched Cepheids, in the reddening-independent Wesenheit index $W_{I}=I-1.55(V-I)$. We find that A-type Cepheids have $P<5.3$ days, while F-type Cepheids span the whole range of periods as expected, when considering the shape of the instability strip on the Hertzsprung -- Russell (HR) diagram, which confines the A-type luminosity range to lower mass stars, while F-type stars span a wider luminosity range. 

\subsubsection {Other periodic stars}
\label{sec:other}
\begin{figure*}
\centering 
\includegraphics[width=6.5in]{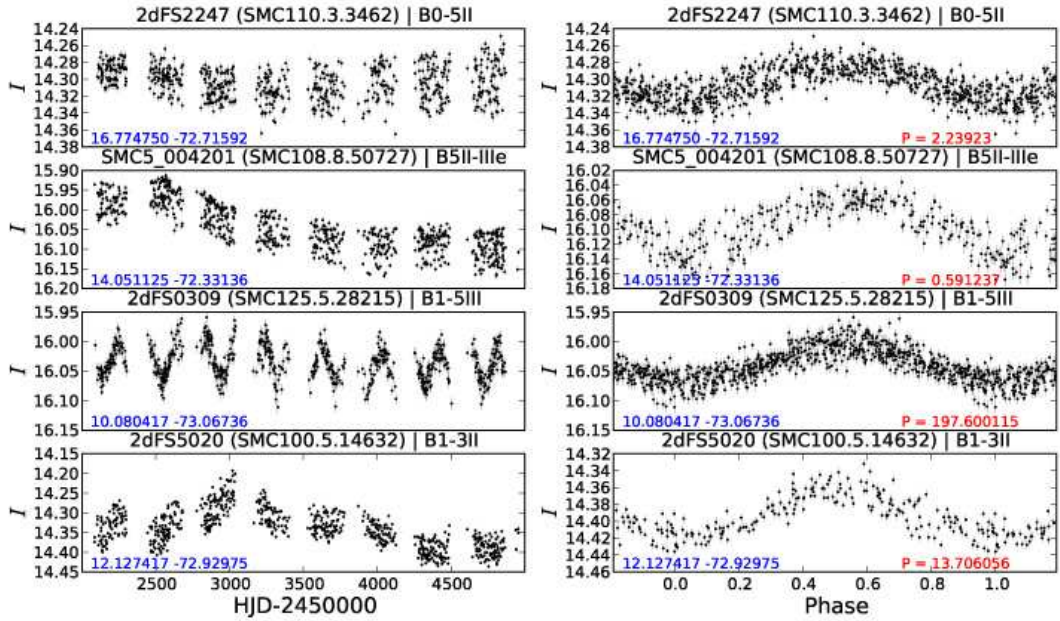}
\caption{$I-$band light curves (left) of 4 periodic variables flagged as ``short'', ``short with extra variability'', ``long',' and ``long with extra variability'', respectively. These are folded to their derived periods (right). In this sample, the folded light curves of SMC5\_004201 and 2dFS5020, which exhibit extra variability, correspond to the HJD ranges 2454000$-$2455000 and 2454200$-$2455000, respectively.}
\label{fig08}
\end{figure*}

The remaining 249 periodic stars that we detected present mostly sinusoidal behavior and correspond to pulsating stars, rotating variables with starspot modulations, or ellipsoidal variables with equal minima. We classified these stars as ``short'' and ``long period'' variables, when presenting periods shorter and longer than three days. Moreover, we flagged variables as ``periodic with extra variability'' when their periodic signal varies in amplitude or is modulated with a baseline variability of $\Delta I$ $\ga$ 0.05 mag. In Tables~\ref{tab05} and \ref{tab06}, we present the main characteristics of the 138 short-periodic and 50 short-periodic stars with extra variability, respectively. Similarly, Tables~\ref{tab07} and \ref{tab08} list the characteristics of the 45 long periodic and 16 long periodic with extra variability. Additionally, we note whether the star has been reported as an eclipsing binary in the literature. We have also matched eight long-period variables with the ``Type-3'' stars defined by \cite{Menn02}. By subtracting the long-periodic modulation, we found four stars\footnote{2dFS1126, 2dFS0309, 2dFS0435, 2dFS0096} that exhibit short-term periods that are equivalent to the expected orbital periods of double-periodic variables, as discussed in Sect. \ref{sec:ecl}. In Fig.~\ref{fig08} we show four representative light curves of variables with short and long periodicity, one from each category described above. 

\subsection {Irregular variables}
\label{sec:irregvar}
\begin{figure*} 
\centering
\includegraphics[width=7in]{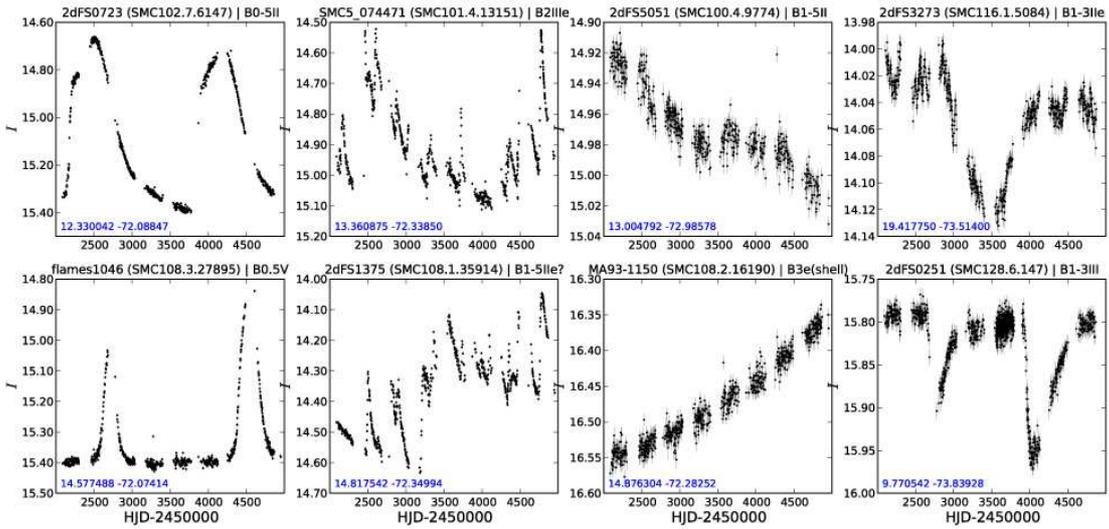}
\caption{Representative pairs of light curves of 8 irregularly varying stars presenting (from left to right) bumper, flicker, monotonic, and fading events.}
\label{fig09}
\end{figure*}

\begin{figure*} 
\centering
\includegraphics[width=7in]{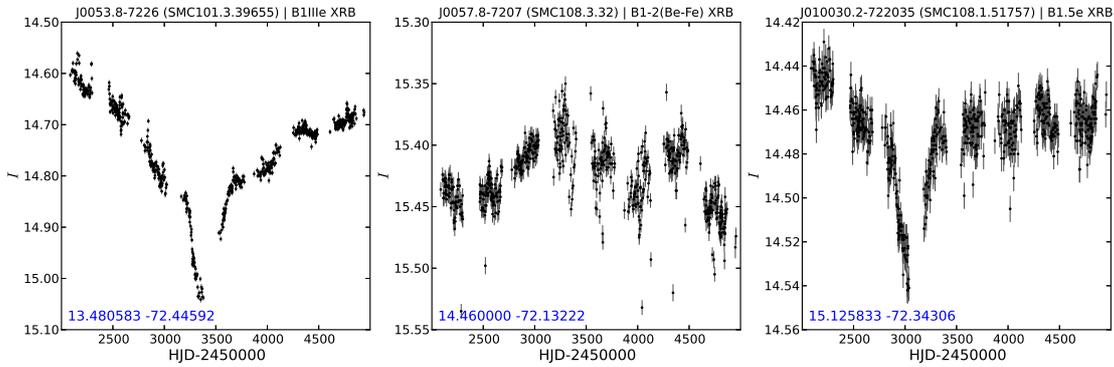}
\caption{Light curves of 3 Be/X-Ray binaries presented for the first time and classified as irregular variables. The first and third exhibit fading events.}
\label{fig10}
\end{figure*}

The morphology of the light curves of the high-amplitude variables, as defined from the $\sigma_{I}$ vs. $I$ plane, can be described as a combination of irregular, short-, or long-term outbursts with amplitudes up to $\Delta I$ $\sim$ 1 mag and/or trends with time scales of hundreds to thousands of days. Based on the classification presented in \cite{Keller02} for blue variables in the LMC, we (re)defined four modes of variability for the $I-$band as follows: 
\begin{itemize}
\item Bumper events - increase in magnitude from the baseline value up to $\Delta I$ $\sim$ 1 mag within a time range of $100-1500$ days;
\item Flicker events - short-term outbursts with an amplitude \\ $\Delta I$ $\gtrsim$ 0.05 mag, usually followed by exponential decay, and a duration of tens of days or even longer for larger amplitude outbursts;
\item Monotonic events - long-term trend across the OGLE-III time domain, with an amplitude $\Delta I$ $\gtrsim$ 0.05 mag;
\item Fading events - decrease in magnitude up to $\Delta I$ $\sim$ 0.6 mag and duration up to 1000 days.
\end{itemize}
The Be phenomenon is known to be enhanced at low metallicities causing outbursts to last longer and have larger amplitudes than in the LMC \citep{Sabogal05}. The bumpers and the fading events reported in the LMC last up to 800 and 600 days, respectively. Variables with this behavior in our sample are found to have outbursts exceeding 1000 days in some cases. In addition, the $V-$band amplitudes exceed --although not significantly-- those reported in the LMC for the same type of photometric variability. Figure~\ref{fig09} presents eight stars with typical bumper, flicker, monotonic, and fading events. Bumper and flicker modes are equivalent to the type-1 hump-like and sharp outbursts, respectively, as defined by \cite{Menn02}. In total, 443 stars were flagged as ``irregular'' and are listed in Table~\ref{tab09}. In the last column we note light curves dominated by a single mode. We cross-matched our irregular variables in the VizieR catalog and also against the photometric study in the SMC by \cite{Menn02}, finding $\sim$76\% of our irregular variables to be newly discovered variables. Furthermore, 23 Be/X-ray binaries compiled by \cite{Raguzova05} have been identified as having irregular variability, four of which present fading events, and one is a bumper variable in the OGLE-III time domain. Of these 23 systems, 20 have been previously studied photometrically. The OGLE-III light curves of the remaining three systems are displayed for the first time in Fig.~\ref{fig10}. 

\begin{figure*} 
\centering
\includegraphics[width=7.5in]{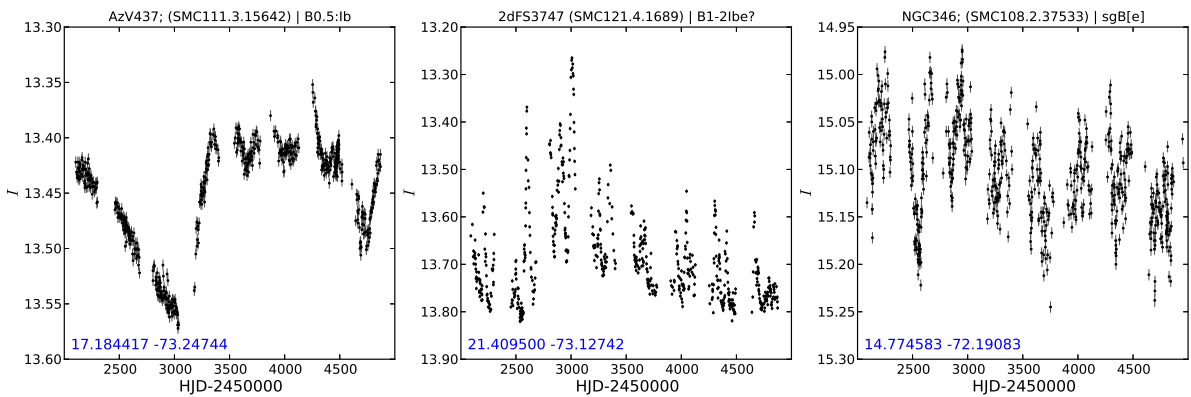}
\caption{Light curves of supergiants presenting ``Be behavior'', including a known sgB[e] star \citep[right panel;][]{Wisniewski07}. Either the luminosity classification is incorrect or their variability is triggered by mechanisms other than fast rotation, such as magnetic fields along with g-mode non-radial pulsations.}
\label{fig11}
\end{figure*}

We have also flagged 21 O, B, and A-type supergiants as irregular since their dispersion places them above the variability threshold on the $\sigma_{I}$ vs. $I$ diagram. Through visual inspection, we conclude that for many of these stars, this is just photometric noise since they are bright stars close to saturation. However, ten of them do exhibit ``Be-like behavior''. Furthermore, six of them have been reported as having emission lines in their spectra \citep[although H$\alpha$ emission could be associated with mass loss, see][]{Leitherer88}, while one is classified as a B[e] supergiant with a circumstellar envelope \citep{Wisniewski07}, and its light curve is displayed in Fig.~\ref{fig11}. By definition, Be stars are not supergiant stars \citep{Collins87}, so we caution that either the luminosity classes of these irregular supergiants may have been overestimated or a mechanism other than fast rotation must be invoked, such as dipole magnetic fields along with non-radial g-mode pulsations \citep{Markova08,Saio11}, which have been reported in the B supergiant region \citep{Saio06,Lefever07,Moravveji12}. 

Table~\ref{tab10} presents the light curves of all irregular and periodic variables discussed in the previous sections and is available in its entirety in the electronic version of the paper and from the CDS.

\subsection {Constant stars}
\label{sec:const}
A total of 2766 stars were found below the threshold of variability as defined by the empirical curve in the $\sigma_{I}$ vs. $I$ diagram in Fig.~\ref{fig02}. Of these, 109 are O-type stars, 1517 early B-type, 347 late B-type, and 793 are AFG-type stars. In general, their light curves appear constant, although some present evidence for relatively low-amplitude variability. However, this is insufficient to establish them as variables with low amplitude using the criterion of deviation. In the present work, we did not search for periodic signals in the region of constant stars, so we expect that further study will identify pulsating variables with low amplitude ($\Delta I$ $\la$ 0.02 mag) and eclipsing binaries with short or shallow eclipses among these stars.

\section{Discussion}
\label{sec:disc}

\begin{figure*}
\centering 
\includegraphics[width=4.5in]{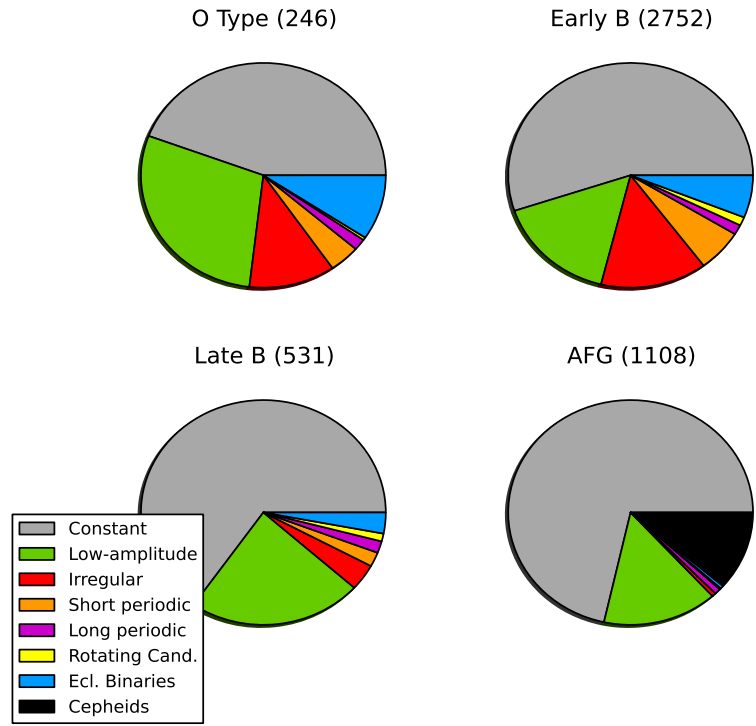}
\caption{Pie charts showing the distribution of the type of variability for the O, early B, late B, and AFG-type stars in our catalog. Numbers in parentheses denote the total number of stars for each spectral type. The frequency of variables --particularly the irregular type-- increases at earlier spectral types. The exact fractions are presented in Table~\ref{tab11} for clarity.}
\label{fig12}
\end{figure*}

\begin{figure*}
\centering 
\includegraphics[width=5in]{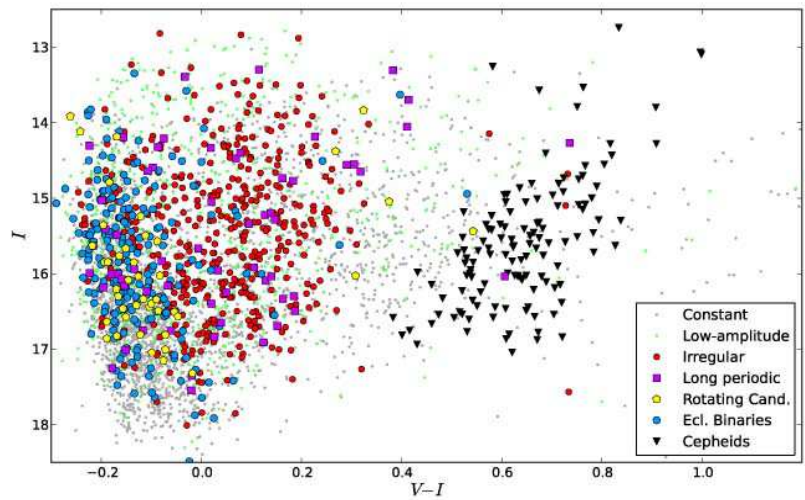}
\caption{Color-magnitude diagram as in Fig.~\ref{fig01}, but for the type of variability. The variability types labeled are constant stars (gray dots), low-amplitude variables (green dots), irregular variables (red circles), eclipsing binaries (blue circles), Cepheids (black triangles), long periodic stars ($P>3$ days; purple squares), and candidate rotating variables (yellow pentagons), as defined in Sect. 4. Short periodic stars do not appear in this diagram for clarity.}
\label{fig13}
\end{figure*}

Having classified 4646 stars by their variability type, we proceeded to investigate their variability as a function of spectral type. Figure~\ref{fig12} presents four pie charts showing the distribution of the seven types of variability (e.g.,\ ``constant'', low-amplitude, irregular, short/long periodic, EBs, and Cepheids) derived through the procedure described in Sect. 4, for 246 O, 2752 early B, 531 late B, and 1108 AFG-type stars. The exact fractions are presented in Table~\ref{tab11}. Nine Wolf Rayet stars do not appear in the statistics, six of which are flagged as low-amplitude variables, two as irregular, and one as a known eclipsing binary. It is striking that variability is so frequent among early-type stars. In fact, since we did not search the ``constant'' region of the $\sigma_{I}$ vs. $I$ diagram for periodicity, we consider the percentage of constant stars to be an upper limit. Similarly, the fraction of $6.1\pm0.5\%$ of early B-type stars that present short-term periodic variability corresponds to a lower limit. Stochastic, high-amplitude variability, as currently reflected in the fraction of irregular variables, peaks at O and early B-type stars, while the decrease beyond the B2 type is considerable. Assuming that irregular variability is mainly caused by the Be phenomenon, we expect it to be associated with an infrared excess and reddened optical colors due to free-free emission originating in the circumstellar region. Indeed, most of the irregular variables exhibit reddening, as shown in Fig.~\ref{fig13}, which presents a color-magnitude diagram (CMD) labeling the variables classified in this work. The instability strip of the Cepheids is distinctive, although it hosts a few other variables, which were examined and excluded from being Cepheids. Of these, four resemble RV Tauri variables considering the morphology of their light curves, and they obey the typical PL relation for Type II Cepheids. Nonetheless, they have low amplitudes and eventually, two of these were classified as long-period variables, one as a candidate rotating variable, and one as an eclipsing binary. The ``constant'' stars and the low-amplitude variables populate the background of the CMD and comprise $\sim$77\% of our studied stars. 

We propose irregular and low-amplitude variability for OB-type stars to be a criterion for defining a candidate photometric Be star. If we apply this criterion to our catalog, then the fraction of candidate early Be stars, i.e.\ early-type stars with irregular or low-amplitude variability, is found to be $30\pm1\%$. This  agrees with the fraction $26\pm4\%$ of Be stars obtained in the spectroscopic study of B-type stars by \cite{Martayan07} in the surrounding field of the open cluster NGC~330 and with the fraction $27\pm2\%$ of Be stars derived by \cite{Bonanos10} using mid-infrared photometry of early B-type stars. This result is also consistent with the study of four young SMC clusters by \cite{Wisniewski06} with H$\alpha$ photometry that yielded a fraction $32\pm5\%$ of candidate Be among early B-type stars. We caution that our fraction of Be stars would increase if we included short periodic stars such as Be pulsating stars, but would also decrease because some irregular and variables with low amplitude may not vary due to the Be mechanism. On the other hand, O-type stars exhibit strong stellar winds driven by radiation pressure that are likely to present low-amplitude variability. After locating our 71 O-type low-amplitude variables on the CMD diagram, only six show evidence of infrared excess. Moreover, out of our ten spectroscopically identified Oe stars, only one is flagged as a low-amplitude variable. We therefore do not find low-amplitude variability to be a signature of the Oe phenomenon. We considered our 28 irregular O-type variables, excluding two that are probably close to saturation and adding the six ``redder'' variables with low amplitude, and found the fraction of candidate Oe stars to be $13\pm2\%$, which is consistent with the fraction $10\pm2\%$ obtained in the mid-infrared study of \cite{Bonanos10}. The agreement of these different methods provides evidence of the validity of the proposed method based on photometric variability. 

\begin{figure*}
\centering
\includegraphics[width=6in]{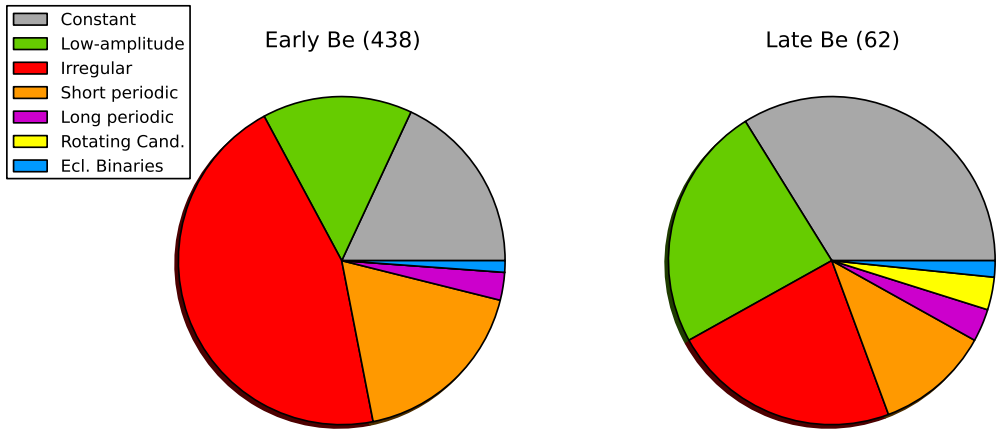}
\caption{Pie charts for 438 early B and 62 late B-type stars that are confirmed spectroscopically as Be stars, showing the distribution of variability types across the 8-year time domain of OGLE-III. We find $78\%$ of the early Be stars and $58\%$ of late Be stars to vary photometrically within the photometric precision of OGLE-III, as a result of disk modulation and/or short-term pulsations. The exact fractions are presented in Table~\ref{tab11} for clarity.}
\label{fig14}
\end{figure*}

Figure \ref{fig14} presents pie charts showing the photometric behavior for the spectroscopically confirmed Be stars with early B and late B types. The exact fractions are presented in Table~\ref{tab11}. We find that $45\pm3\%$ of early-Be stars present irregular variability. Adding our stars with stochastic low-amplitude variability, as well as the pulsating Be variables, the fraction of photometrically variable Be stars increases to $78\pm4\%$. In addition, since part of the spectroscopic observations of the Be stars were conducted before the OGLE-III monitoring period, we recalculated the statistics using only the 128 early-Be stars from \cite{Evans06} and \cite{Martayan07} that conducted spectroscopy during the OGLE-III monitoring. Such a selection yields a consistent fraction of $79\pm8\%$. On the other hand, the backbone of the catalog comes from \cite{Evans04} and corresponds to $78\%$ of the total sample. The spectra were obtained in 1998 and 1999, yielding 228 of our 438 studied early-Be stars. Using only the Be stars from this survey, the fraction of Be stars that exhibit photometric variability from the subsequent OGLE-III light curves is $79\pm6\%$. We caution, though, that our pulsating fraction is a lower limit since it is incomplete and that the fraction of low-amplitude variable Be stars could be higher, considering that most of them exhibit long-lasting invariant intervals; therefore, when conducting spectroscopy, they have a lower chance of being detected with H$\alpha$ emission than irregular variables. In addition, \cite{Evans04} denoted 46 early B-type stars of our sample to have an uncertain (``e?'') spectral classification. Of these, one third were classified in our study as ``constant'', while the respective fraction of ``constant'' stars among the typical ``e'' stars is one sixth. Therefore, stars with controversial emission, e.g.\ nebular emission, might have contaminated our Be sample and appear as ``constant',' although we caution that some of these could indeed vary relatively insignificantly. We therefore report a minimum fraction $\sim$$80\%$ of early B-type stars displaying the Be phenomenon that present significant photometric variability on a eight-year timescale. This percentage is comparable to $86\%$ provided by \cite{Hubert98} for short-term variability of Be stars in the Galaxy. The fraction of their short-term period variables in the B0-B3 interval is $\sim$$40\%$, much higher than our incomplete $18\%,$ although the expected fraction $\sim$$25\%$ for pulsating Be stars in the SMC \citep{Diago08} is still lower than for the Galaxy, indicating the dependence of the pulsations on metallicity \citep{Pam99}.
 
It is remarkable that $\sim$53\% of our candidate photometric early Be stars, i.e.\ low-amplitude and irregular early B-types, are stars of the luminosity class range I-III. As already mentioned, winds and saturation noise are possibly the cause of the low-amplitude variability that is prominent in some of our brightest stars. However, we caution that the presence of a circumstellar envelope causes the Be stars to appear overluminous, while rapid rotation could modify their fundamental parameters by shifting the effective temperature or by leading to an underestimation of v$\sin i$ due to the gravitational darkening effect \citep{Townsend04}. As a result, a fast-rotating star could appear cooler and more luminous. \cite{Martayan07} implemented fast rotation treatment for their spectroscopic observations in the region of NGC~330, which decreased the fraction of giant Be stars from $\sim$50\% to $\sim$10\%. The main sequence Be fraction increased from $\sim$15\% to $\sim$40\%, while the fraction of subgiant field Be stars remained at $\sim$45\%.

\begin{figure*} 
\centering
\includegraphics[width=7in]{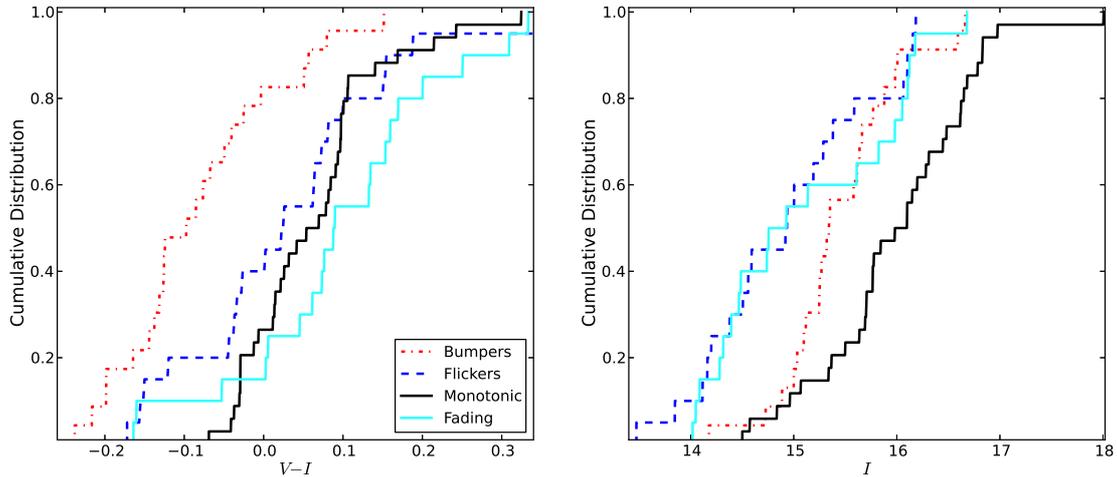}
\caption{Cumulative distribution diagrams of stars that present a single mode of irregular variability, as a function of their $V-I$ color (left panel) and their $I-$band magnitude (right panel). Bumper stars are shown with a red dashed-dotted line, flicker types with a blue dashed line, monotonic types with a black solid line, and fading types with a cyan solid line. We note that all four modes exhibit colors between $-$0.2 and 0.3 mag. The colors of bumper stars are shifted towards the blue only because of their insufficient $V-$band monitoring. Flicker-type stars appear to be the brightest among the modes. About half of the single-mode irregular variables are spectroscopically confirmed as Be stars.}
\label{fig15}
\end{figure*}

All four modes of irregular variability exhibit a range of $V-I$ colors between $-$0.2 and 0.3 mag as shown in the cumulative distribution diagram for the average $V-I$ color of the light curves in the lefthand panel of Fig.~\ref{fig15} indicating a range of reddening, which is probably due to the size of the circumstellar disk and its lifetime. The mean color for our bumper type stars is shifted towards the blue because a significant part of their $V-$band photometry was conducted at intervals when they behaved as regular B stars, i.e.\ before or after the outbursts, and thus, their color was mainly photospheric. Stars with monotonic behavior seem to be fainter than those exhibiting outbursts, as shown in the cumulative distribution diagram for $I-$magnitudes in the righthand panel of Fig.~\ref{fig15}, while flicker type stars appear brighter than the other modes. Of our single-mode irregular variables, $\sim$53\% are spectroscopically confirmed Be stars.  

We identified 11 single-variability mode stars from our catalog as among the Be stars for which \cite{Martayan07} measured parameters: one bumper, three flicker, five monotonic, and two fading types. The bumper variable SMC5\_078440 corresponds to a fast rotator with corrected velocity v$\sin i_{true} \sim$ $380\pm20$ km~s$^{-1}$. Furthermore, for the three flicker-type stars\footnote{SMC5\_074471, SMC5\_021152 and SMC5\_073581}, the mean obtained velocity is v$\sin i_{true} \sim$ $220\pm10$ km~s$^{-1}$, and for the five stars with monotonic behavior\footnote{SMC5\_078338, SMC5\_075360, SMC5\_037137, SMC5\_003919 and SMC5\_064745} the respective value is v$\sin i_{true} \sim$ $290\pm10$ km~s$^{-1}$. Finally, the two fading type variables SMC5\_046388 and SMC5\_065055 exhibit a true mean velocity v$\sin i_{true} \sim$ $400\pm30$ km~s$^{-1}$. Stars with flicker type outbursts seem to be average rotators, and therefore, their parameters are less affected by the rotation. As a result, we suggest that their luminosity is intrinsic rather than due to fast rotation. We suspect that the flickering behavior may occur mostly in evolved stars, although we need a large dataset to confirm this.

\begin{figure*} 
\centering
\includegraphics[width=6.5in]{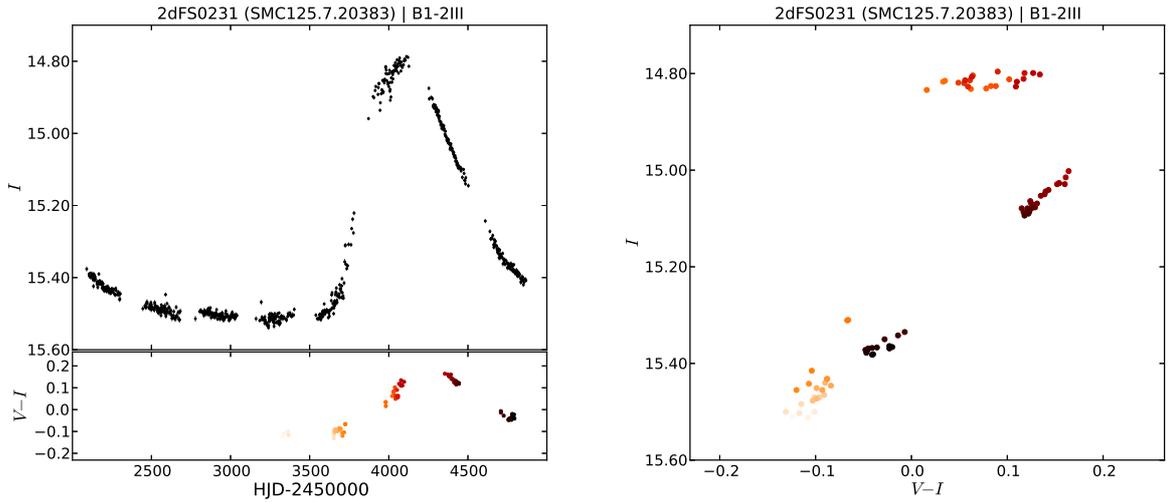}
\caption{The left panel shows the $I-$band light curve (upper) and $V-I$ color modulation (lower) of a bumper event exhibiting clock-wise loop structure in the color-magnitude diagram, indicating outflowing material that accumulates in a disk that transits from an optically thick to optically thin state \citep{deWit06}. The right panel shows the color-magnitude diagram where the loop structure is demonstrated.}
\label{fig16}
\end{figure*}

\citet{deWit06} have proposed a model for outbursts describing the establishment and dissipation of the disk and subsequent modifications of the optical depth. These authors predict that the outflowing disk material produces a clock-wise, loop-like structure (vs.\ counter clock-wise loop for the rare case of inflowing material) in the color-magnitude diagram. This bi-valued relation between color and magnitude is produced by the transition of the disk from an optically thick to thin state. Of our 23 bumper variables, we find five stars exhibiting a full-loop, clock-wise structure (Fig.~\ref{fig16}) with mean color amplitude $\Delta (V-I)$ $\sim$ 0.35 mag and 12 stars presenting very narrow loops or half-loop evidence due to limited $V-$band photometry. The remaining six bumper variables were monitored in the $V-$band mostly before or after the outburst. On the other hand, a single-valued relation is most prominent in our monotonic type stars, denoting that ejected matter stays optically thin. Flicker type outbursts have a short duration, and because our $V-$band photometry is sparse, we do not have a clear picture of the CMD pattern or the underlying mechanism. Nonetheless, we did detect clock-wise cyclic loops in the flicker-type stars AzV436/2dFS2413 and SMC5\_074471 that have spectral types B0IIe and B2IIIe, respectively. 

\subsection {Short periodic stars}
\label{sec:discsps}
\begin{figure*}
\centering 
\includegraphics[width=6.5in]{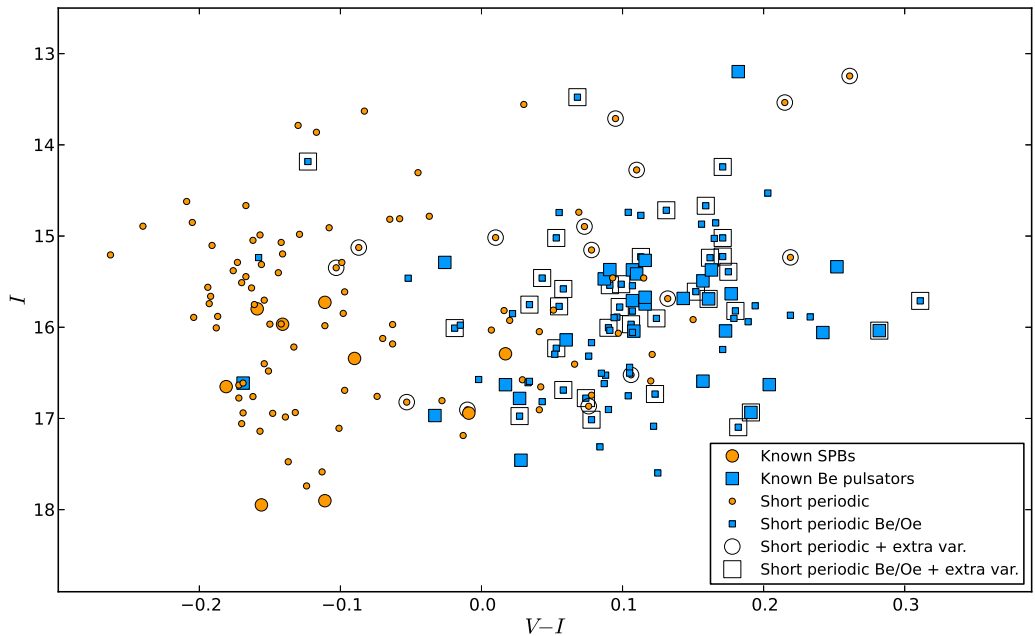}
\caption{Color-magnitude diagram for the short periodic stars ($P<3$ days) in this paper along with the known SPBs and non-radial Be pulsating stars from \cite{Diago08}. Several early B-type stars are likely to be newly discovered SPBs. Also, short-periodic stars with additional variability exhibit redder colors, indicating the origin of such behavior to the Be phenomenon.}
\label{fig17}
\end{figure*}

The reason for selecting the limit of three days to distinguish short-period from long-period variables was to differentiate a class of radial pulsating B-type variables with $P < 3$ days, known as slowly pulsating B stars \citep{deCat02}. Furthermore, non-radial Be pulsators present a common range of periods with SPBs and share the same region in the HR diagram. SPBs are expected to be rare at the low metallicity of the SMC according to computations presented in \cite{Miglio07}. We therefore expect that the discovery of new candidate SPBs could redefine the borders of their instability domain on the HR diagram and update the properties of the metal opacity bump \citep{Salmon12}. In Fig.~\ref{fig17}, we present a CMD comparing the SPBs and the Be pulsators reported by \cite{Diago08} in the SMC that are also found in our catalog among the short periodic stars. Of the 188 short periodic stars, 10 are O-type, 167 are early B-type, and 11 are late B-type stars. Several short period non-Be (or not defined) stars are colocated with known SPBs and, thus, are candidate SPBs. Similarly, many short period Be stars are likely to exhibit non-radial pulsations. We also note that the short periodic stars that show additional variability are redder and consistent with the reported Be pulsators, independent of whether they are found to have Balmer emission. In addition, four Be stars with additional variability studied by \cite{Martayan07} exhibit mean corrected projected velocities v$\sin i_{true}= 490\pm30$ km~s$^{-1}$, while for 15 short periodic Be pulsators without additional variability, the respective value is v$\sin i_{true}= 350\pm10$ km~s$^{-1}$. Therefore, we suggest that such evident fluctuations in the light curves are linked to the Be phenomenon and enhanced by fast rotation. Their g-mode non-radial pulsations could be excited by the subsurface iron convection zone, although it is expected to be less prominent at the metallicity of the SMC \citep{Cantiello09}.

The $\beta$ Cephei stars are another group of blue, short-period, pulsating variables. These are not expected at the metallicity of the SMC; however, several candidate $\beta$ Cephei stars have been reported in the SMC \citep{Kolac06,Sarro09,Diago08}, and their presence possibly denotes metal-rich stars in the galaxy but could also denote an uncertainty in the opacity of nickel by a factor of two \citep{Salmon12}. Among our short periodic stars, we have identified seven candidate $\beta$ Cephei stars. These are early-type, multiperiodic variables with blue colors ($V-I<0$) that present low-amplitude sinusoidal periodicity up to $0.5$ days. We flag these in Table~\ref{tab05} and provide the two periods that correspond to the highest frequency peaks.

\subsection {Eclipsing binaries}
\label{sec:discecl}
\begin{figure*} 
\centering
\includegraphics[width=6.5in]{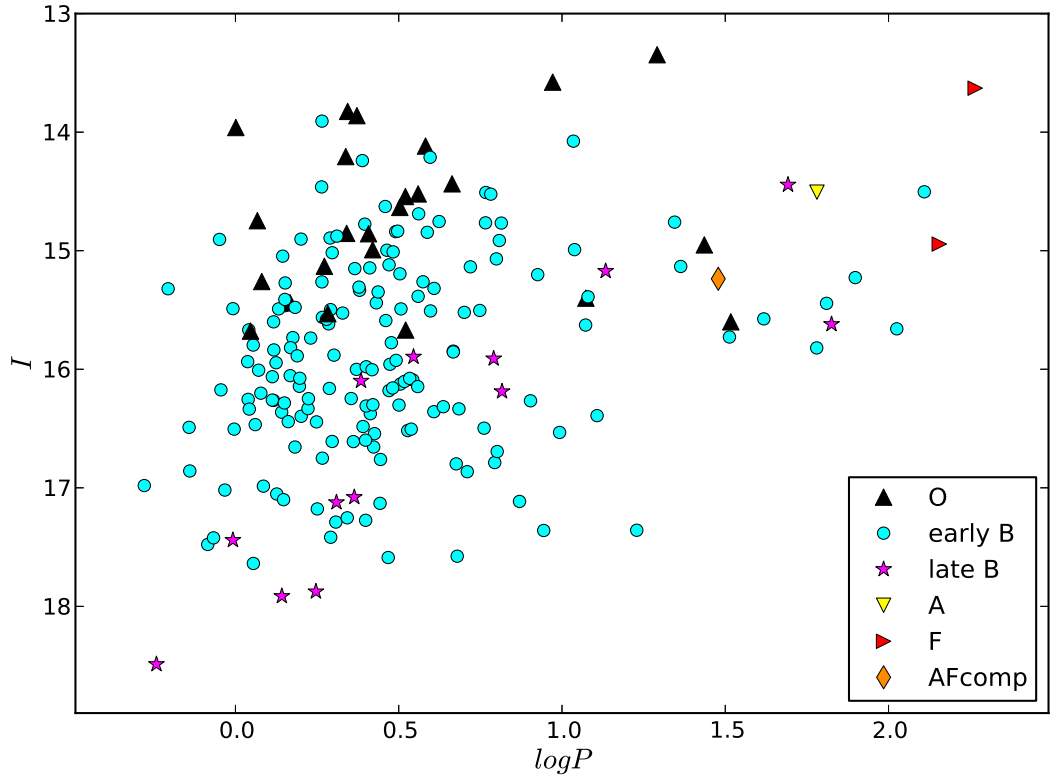}
\caption{Period-magnitude diagram for the known and the newly discovered eclipsing binaries of our sample, showing a prominent concentration of early type EBs in short period orbits and a distinctive tail of luminous EBs towards longer periods. All four AFG-type EBs exhibit periods over $\sim$30 days.}
\label{fig18}
\end{figure*}

Through the present work, we doubled the number of the massive EBs known in the SMC from 104 to 205. The vast majority of them lie close to the main sequence. Figure~\ref{fig18} presents their period-magnitude relation as a function of  their spectral type. The majority of O-type binaries are short-period systems, while B-type systems reach an upper limit of about 130 days. On the other hand, the four A and F-type eclipsing binaries identified have long periods. Two of these, 2dFS0142 and SMC3-4 \citep{Menn10}, are supergiants with wide orbits, and one contains composite features of A and F-type stars, as discussed in Sect. \ref{sec:ecl}. The newly discovered F0 type EB 2dFS2348, along with the likely ellipsoidal variable 2dFS0951 of type A7Ib reported by \cite{Wyrz04} exhibit similar colors $V-I\sim0.5$. The morphology of their light curves reveals they are contact binaries, while the unequal eclipse depths suggest systems with one evolved component overflowing its Roche lobe along with a bluer companion, which shifts the mean color. We proceeded to compare these to the yellow SMC supergiant R47 \citep{Prieto08} of similar color, with absolute magnitude $M_{V}=-7.5$, which is identified as a likely progenitor of Type IIP supernova. Nonetheless, our systems have rather low amplitude, and their $V-$band magnitudes indicate less massive components with $M_{V}\sim-3.5$.

Short-period binaries with closer orbits containing large, massive stars, as in the case of O-type systems, are more likely to be detected in eclipsing configurations. They therefore dominate our eclipsing sample. To evaluate the fraction of eclipsing binaries in each spectral bin, we eliminated possible selection effects of the compiled input catalog by removing the eclipsing binaries studied by \citet{Hild05} and \citet{Harries03}, as well as additional selected X-ray binaries and periodic variables. The observed fractions of eclipsing systems are $5.2\pm1.5\%$ for O-type, $4.8\pm0.4\%$ for early B-type, $2.7\pm0.7\%$ for late B-type, and unsurprisingly, only $0.3\pm0.2\%$ for AFG-type stars since evolved and luminous stars mainly belong to wide, long-period systems \citep{Wyrz04,Udalski98} and, therefore, are rarely identified.
 
\section{Conclusions}
We used the photometry provided by the OGLE-III monitoring survey to study the variability of 4646 massive stars with known spectral types in the SMC. This study of the variability of massive stars with known spectral types is a factor of 7 larger than the work of \cite{Szczygiel10}. A total of 2766 stars that correspond to $\sim$60\% of our sample present invariant light curves given the accuracy of the photometry. We detected stochastic low-amplitude variability with a mean $\Delta I$ $\la$ 0.05 in 807 stars and variability of higher amplitude in 443 stars. We also report the discovery of 101 EBs\footnote{These EBs became part of the recently released OGLE-III set of EBs in the SMC \citep{Pawl13}.} with massive components and present the OGLE-III photometry for 104 known EBs. We hope these will motivate follow-up studies to determine their fundamental parameters and contribute to the small sample of massive stars with accurately measured parameters. Moreover, we have detected short-term periodicity ($P < 3$ days) in 188 stars, thereby motivating studies on new candidate SPBs and non-radially pulsating Be variables. Having inspected four pulsating Be stars with fluctuating amplitude or baseline, we find that they are rapid rotators close to their critical speed, suggesting that fast rotation is responsible for their additional variability most likely caused by matter ejections. 

Long-term photometry is established as a powerful tool for studying massive stars, determining accurate periods for known EBs, and discovering rare long-period systems. It provides a new method for discovering new candidate Be stars that were not detected before, either because they were not subject to H$\alpha$ spectroscopy or they were observed during their transition back to their B state owing to the dissipation of their circumstellar gas disk. A fraction of $51\pm4\%$ of our irregular early B stars have been already identified spectroscopically as Be stars. The remaining irregulars ($\sim$190 stars) therefore comprise new candidate Be stars. Furthermore, $\sim$20 O-type stars that exhibit irregular variability are likely new candidate Oe stars. On the other hand, the fraction of low-amplitude early B-type variables identified as Be stars is only $15\pm2\%$ mainly because their sparse and/or short-term fluctuations make their spectroscopic identification as Be stars unlikely. We note, though, that not all low-amplitude variables are expected to be linked to the Be phenomenon.

This work is just a foretaste of the results that LSST \citep{Ivezic08} and Pan-STARRS \citep{Magnier13} will produce for the nearby galaxies that they will survey. Similar follow-up research is possible to be carried out on the LMC, based on the catalog of \cite{Bonanos09b}, which consists of massive stars with known spectral type, with the goal of studying and comparing the Be phenomenon from the photometric perspective among environments of different metallicity.

\begin{acknowledgements}
We thank the anonymous referee and R. Mennickent for helpful comments that have improved the manuscript. M. Kourniotis and A.Z. Bonanos acknowledge research and travel support from the European Commission Framework Program Seven under the Marie Curie International Reintegration Grant PIRG04-GA-2008-239335. The OGLE project has received funding from the European Research Council under the European Community's Seventh Framework Program (FP7/2007-2013)/ERC grant agreement no. 246678. This research made use of the VizieR catalog access tool at the CDS, Strasbourg, France, and of NASA's Astrophysics Data System Bibliographic Services. 
\end{acknowledgements}


\begin{sidewaystable*}

{\tiny
\caption{New eclipsing binaries in the SMC}
\label{tab01}
{\centering
\begin{tabular}{lllccrrlrrrc}
\hline \hline
\# & Star name & Spectral type & R.A. (J2000)   & Dec. (J2000)    & Skewness & Kurtosis & OGLE-III ID & $<$I$>$ & $<$V$>$ & Period (days) & Note \\  
\hline
91  & 2dFS3361                                 & B2IV                       & 19.743875 & $-$72.75589 &  0.274 &$-$0.136 &   SMC120.7.4952  & 16.09 & 15.95 &   3.492387 &   \\
92  & flames1175;SMC5\_027763                  & B2V                        & 14.334471 & $-$72.21558 &  0.484 &$-$0.216 &   SMC108.7.49913 & 16.46 & 16.31 &   1.149758 &   \\  
93  & flames1014;SMC5\_006463                  & B3Ib                       & 14.988021 & $-$72.13126 &  1.784 &  3.996  &   SMC108.3.14365 & 14.44 & 14.38 &  49.174040 &  SB*    \\ 
94  & 2dFS3495                                 & B3-5III                    & 20.258375 & $-$72.34086 &  0.647 &$-$0.685 &   SMC123.1.32    & 16.09 & 15.99 &   2.423943 &   \\
95  & NGC330-079;SMC5\_037369                  & B3III                      & 13.968375 & $-$72.43648 &  0.311 &  0.088  &   SMC108.8.17533 & 16.18 & 16.07 &   6.551927 &   \\
96  & 2dFS2950                                 & B8-A0III                   & 18.384167 & $-$73.89856 &  0.515 &$-$0.442 &   SMC117.6.5502  & 17.91 & 17.93 &   1.387038 &   \\
97  & 2dFS2838                                 & B8:III                     & 18.083583 & $-$73.47872 &  1.725 &  2.421  &   SMC116.7.9732  & 17.12 & 17.06 &   2.038393 &   \\
98  & 2dFS1036                                 & B8IIe?                     & 13.747708 & $-$72.14592 &  0.739 &$-$0.160 &   SMC108.6.5879  & 15.17 & 15.24 &  13.604933 &   \\
99  & 2dFS0275                                 & B9:III                     &  9.908833 & $-$73.89717 &  0.924 &  0.018  &   SMC128.6.1180  & 17.44 & 17.45 &   0.982208 &   \\
100 & AzV465                                   & Bpec                       & 17.957500 & $-$72.73558 &  0.713 &  1.827  &   SMC115.6.10187 & 15.20 & 15.01 &   8.423822 &    AM \\
\hline
\end{tabular}   

\tablefoot{Table~\ref{tab01} is available in its entirety in the electronic version of the Journal. A portion is shown here for guidance
regarding its form and content. \\ \textbf{Keys:} (AM) Apsidal Motion, (TEB) Transient EB, (DPV) Double-Periodic Variable, (*)~Hunter et al. (in preparation), (**)~\cite{Evans06} \\ $^a$Period corresponds to HJD range 2452000$-$2454200}

} 

} 

\end{sidewaystable*} 


\begin{sidewaystable*}

{\tiny
\caption[]{Known eclipsing binaries in the SMC}

\label{tab02} 
{\centering
\begin{tabular}{lllccrrlrrrrc}

\hline \hline
\# & Star name & Spectral type & R.A. (J2000)   & Dec. (J2000)    & Skewness & Kurtosis & OGLE-III ID & $<$I$>$ & $<$V$>$ & Reference & Period (days) & Note \\  
\hline
43  & flames1101;SMC5\_028846              & B0.5:             & 14.812408 & $-$72.19310 & 2.998 &   10.102 &  SMC108.2.37672 & 15.92 & 15.78 & W04 &   3.104372 &   AM  \\
44  & flames1097;SMC5\_006928              & B0.5:+early B     & 14.601738 & $-$72.09236 & 1.891 &    2.722 &  SMC108.3.109   & 16.05 & 15.81 & S07 &   1.471118 &  SB2* \\
45  & AzV204;2dFS1305                      & B0.5III           & 14.591625 & $-$72.58792 & 0.649 & $-$0.437 &  SMC105.4.16    & 14.50 & 14.47 & W04 &   5.844730 &     \\
46  & 2dFS2283                             & B0.5IV            & 16.839792 & $-$72.20733 & 1.338 &    1.370 &  SMC113.2.24785 & 14.87 & 14.60 & F07 &   2.044898 &     \\
47  & 2dFS0757                             & B0.5V             & 12.577125 & $-$72.64892 & 0.976 & $-$0.312 &  SMC101.2.20    & 15.26 & 15.10 & F07 &   1.839888 &     \\
48  & NGC330-030;SMC5\_020391              & B0.5V             & 14.098250 & $-$72.35657 & 1.644 &    1.566 &  SMC108.8.50687 & 15.15 & 14.99 & F07 &   2.320057 &  SB** \\
49  & NGC330-055;SMC5\_012510              & B0.5V             & 14.205167 & $-$72.49138 & 0.519 &    1.002 &  SMC105.5.37411 & 15.85 & 15.67 & W04 &   4.646085 &     \\
50  & 2dFS2090                             & B0.5V             & 16.490167 & $-$72.19814 & 0.615 &    0.546 &  SMC113.2.16414 & 15.27 & 15.01 & F07 &   1.420025 &   +TEB$^{a}$  \\
51  & AzV44                                & B0IIWW            & 12.197625 & $-$73.41650 & 0.913 & $-$0.412 &  SMC100.8.14628 & 13.90 & 13.68 & W04 &   1.840730 &     \\
52  & 2dFS0498                             & B0IV              & 11.036000 & $-$73.23811 & 2.881 &    7.994 &  SMC125.3.41165 & 14.52 & 14.39 & W04 &   6.051906 &   AM  \\
\hline
\end{tabular} 
 
\tablefoot{Table~\ref{tab02} is available in its entirety in the electronic version of the Journal. A portion is shown here for guidance
regarding its form and content. \\ \textbf{References:} (W04) \citet{Wyrz04}; (F07) \citet{Faccioli07}; (N02) \citet{Niemela02}; (U98) \citet{Udalski98}; (G05) \citet{Groen05}; \\ (H03) \citet{Harries03}; (H05) \citet{Hild05}; (B02) \citet{Bayne02}; (S09) \citet{Samus09}; (M06) \citet{Menn06} \\ \textbf{Keys:} (AM) Apsidal Motion, (TEB) Transient EB, (*)~Hunter et al. (in preparation), (**)~\cite{Evans06} \\ $^a$Additional transient EB with $P=5.242314$ days, in the HJD range 2453200$-$2453800}
 
} 

} 
\end{sidewaystable*} 


\begin{table*}

{\tiny
\caption[]{Rotating variables / Overcontact binaries}
\label{tab03}
{\centering
\begin{tabular}{lllcclrrrc}
\hline \hline
\# & Star name & Spectral type & R.A. (J2000)   & Dec. (J2000) & OGLE-III ID & $<$I$>$ & $<$V$>$ & Period (days) & Note \\ 
\hline
1   & AzV387;2dFS2050                            &  O9.5III                  &  16.416458 &   $-$72.26822 &     SMC113.2.8     &   14.11  &  13.87 &     2.972105  &                            \\ 
2   & 2dFS2698                                   &  B0-5II                   &  17.676917 &   $-$73.32697 &     SMC116.6.129   &   15.85  &  15.71 &     6.405136  &      \\
3   & 2dFS0300                                   &  B0-5IV                   &  10.037958 &   $-$73.64939 &     SMC128.5.15134 &   16.63  &  16.44 &     2.368890  &   W04              \\ 
4   & 2dFS2455                                   &  B0-5IV                   &  17.176375 &   $-$72.32797 &     SMC113.1.31538 &   16.02  &  15.94 &     3.528916  &   W04              \\ 
5   & 2dFS0359                                   &  B0-5IV                   &  10.308958 &   $-$73.42589 &     SMC125.7.14593 &   16.55  &  16.69 &     1.667506  &                      \\
6   & 2dFS0010                                   &  B0-5V                    &   7.349375 &   $-$73.73997 &     SMC138.4.5904  &   17.32  &  17.30 &     3.561880  &                            \\ 
7   & 2dFS1421                                   &  B0-5V                    &  14.956333 &   $-$72.54781 &     SMC105.4.5344  &   16.49  &  16.40 &     1.320202  &                            \\ 
8   & 2dFS2295                                   &  B0-5V                    &  16.875167 &   $-$73.43469 &     SMC111.2.2145  &   16.56  &  16.51 &     4.326558  &                            \\ 
9   & 2dFS2959                                   &  B0-5V                    &  18.426875 &   $-$72.63375 &     SMC115.3.162   &   16.86  &  16.67 &     0.856288  &                            \\ 
10  & 2dFS3304                                   &  B0-5V                    &  19.517333 &   $-$72.65019 &     SMC120.6.146   &   16.80  &  16.68 &     1.062355  &                            \\ 
\hline
\end{tabular} 

\tablefoot{Table~\ref{tab03} is available in its entirety in the electronic version of the Journal. A portion is shown here for guidance regarding its form \\ and content. \\ \textbf{References:} (W04) \citet{Wyrz04}; (D08) \citet{Diago08}; (U98) \citet{Udalski98}; (M10) \citet{Menn10} \\ \textbf{Keys:} (*)~Hunter et al. (in prep.), (**)~\cite{Evans06}, (***)~\cite{Martayan07}}
} 

} 
\end{table*} 


\begin{table*}
{\tiny
\caption[]{Known Cepheids in the SMC}
\label{tab04}  
\begin{tabular}{lllccrrlrrr}
\hline \hline
\# & Star name & Spectral type & R.A. (J2000)   & Dec. (J2000) & Skewness & Kurtosis & OGLE-III ID & $<$I$>$ & $<$V$>$ & Period (days) \\ 
\hline
1  & 2dFS1914           & A0II                 &  16.165833 &$-$72.31958 &$-$0.227 &$-$1.049 & SMC113.8.39669 &  15.45 & 16.12  &  3.35206   \\
2  & 2dFS1472           & A2Iab                &  15.115958 &$-$73.37806 &$-$0.271 &$-$0.976 & SMC106.2.30024 &  16.38 & 17.10  &  2.18004   \\
3  & 2dFS1719           & A2Ib                 &  15.737167 &$-$73.41747 &  0.087  &$-$1.341 & SMC111.7.58    &  15.51 & 16.26  &  5.19790   \\
4  & 2dFS1269           & A3Ib                 &  14.467083 &$-$74.10453 &$-$0.103 &$-$1.429 & SMC107.8.8007  &  16.75 & 17.17  &  1.08680   \\
5  & 2dFS2327           & A3Ib                 &  16.936458 &$-$73.19008 &$-$0.048 &$-$1.400 & SMC111.3.12630 &  16.22 & 16.74  &  1.54276   \\
6  & 2dFS2460           & A3Ib                 &  17.185833 &$-$72.69247 &$-$0.316 &$-$1.027 & SMC110.3.10311 &  15.63 & 16.45  &  3.71093   \\
7  & 2dFS0667           & A3II                 &  11.973833 &$-$72.83844 &  0.124  &$-$1.033 & SMC100.5.37708 &  15.98 & 16.41  &  3.28218   \\
8  & 2dFS0960           & A3II                 &  13.460917 &$-$74.02661 &$-$0.526 &$-$0.817 & SMC103.1.7083  &  16.11 & 16.80  &  2.70204   \\
9  & 2dFS1403           & A3II                 &  14.898833 &$-$73.05503 &$-$0.428 &$-$1.044 & SMC106.4.28330 &  17.04 & 17.66  &  0.68047   \\
10 & 2dFS2353           & A3II                 &  16.990958 &$-$73.30878 &$-$0.462 &$-$0.895 & SMC111.3.3080  &  16.50 & 17.02  &  1.85038   \\
\hline
\end{tabular} 

\tablefoot{Table~\ref{tab04} is available in its entirety in the electronic version of the Journal. A portion is shown here for guidance regarding its form \\ and content.}           

}
\end{table*} 


\begin{table*}
{\tiny
\caption[]{Short-period stars in the SMC}
\label{tab05} 
\begin{tabular}{lllcclrrrc}
\hline \hline
\# & Star name & Spectral type & R.A. (J2000)  & Dec. (J2000) & OGLE-III ID & $<$I$>$ & $<$V$>$ & Period (days) & Note\\ 
\hline
22 & 2dFS1910                                           &  B0-5IV                       &   16.156542 & $-$72.19458   &   SMC113.7.40193  & 16.48  &  16.32   &    0.705250 &           \\
23 & 2dFS2522                                           &  B0-5IV                       &   17.290375 & $-$73.45019   &   SMC111.2.4284   & 16.77  &  16.60   &    0.993426 &           \\
24 & 2dFS3214                                           &  B0-5IV                       &   19.187333 & $-$73.30769   &   SMC116.3.7028   & 16.40  &  16.47   &    0.602205 &           \\
25 & 2dFS3518                                           &  B0-5IV                       &   20.370875 & $-$72.75442   &   SMC120.2.45     & 15.97  &  15.90   &    0.579404 &           \\
26 & 2dFS1551                                           &  B0-5IVe                      &   15.320458 & $-$72.76842   &   SMC105.2.37523  & 16.44  &  17.74   &    0.534116 &     W04   \\    
27 & 2dFS1142                                           &  B0-5IVe?                     &   14.085417 & $-$73.60042   &   SMC106.8.10982  & 16.31  &  16.39   &    1.241582 &           \\
28 & 2dFS1773                                           &  B0-5IVe?                     &   15.874375 & $-$72.97458   &   SMC110.8.4493   & 16.75  &  16.85   &    0.309992 &           \\
29 & 2dFS2164                                           &  B0-5IVe?                     &   16.612167 & $-$72.28303   &   SMC113.2.4478   & 16.16  &  16.24   &    0.656981 &           \\
30 & 2dFS0205                                           &  B0-5V                        &    9.520958 & $-$73.54000   &   SMC125.8.16896  & 17.05  &  16.88   &    0.381540 &  BCEP(+0.275963d)         \\
31 & 2dFS0421                                           &  B0-5V                        &   10.619042 & $-$72.65756   &   SMC126.3.431    & 17.58  &  17.47   &    0.612584 &           \\
\hline
\end{tabular} 

\tablefoot{Table~\ref{tab05} is available in its entirety in the electronic version of the Journal. A portion is shown here for guidance regarding its form \\ and content.\\ \textbf{References:} (W04) \citet{Wyrz04}; (D08) \citet{Diago08}; (I04) \citet{Ita04} \\ \textbf{Key:} (BCEP) Candidate $\beta$ Cephei}     

}
\end{table*} 


\begin{table*}
{\tiny
\caption[]{Short-period stars with extra variability in the SMC}
\label{tab06}
{\centering
\begin{tabular}{lllcclrrrc}
\hline \hline
\# & Star name & Spectral type & R.A. (J2000)  & Dec. (J2000) & OGLE-III ID & $<$I$>$ & $<$V$>$ & Period (days) & Note\\ 
\hline
1  & AzV480;2dFS3047                          &  O4-7Ve                       &  18.729292  & $-$72.36058   &   SMC118.1.5302   &  14.24  &  14.41 &  0.637494  &  \\
2  & 2dFS2553                                 &  O6.5IIf                      &  17.341583  & $-$73.26150   &   SMC111.3.7734   &  15.01  &  15.02 &  0.823163  &  \\
3  & 2dFS3357                                 &  O9.5III-V                    &  19.725167  & $-$73.16056   &   SMC121.5.11     &  14.27  &  14.38 &  0.812373  &  \\
4  & J005517.9-723853                         &  O9.5Ve XRB                   &  13.824583  & $-$72.64806   &   SMC105.6.33072  &  15.71  &  16.02 &  0.650438  &  \\
5  & AzV322                                   &  O9II                         &  15.750042  & $-$72.42761   &   SMC113.8.6165   &  13.71  &  13.80 &  2.586981  &  \\
6  & AzV162                                   &  O9V                          &  13.753292  & $-$72.93822   &   SMC105.8.34986  &  13.53  &  13.75 &  1.598756  &  W04     \\
7  & J0050.7-7316                             &  B0-0.5Ve XRB                 &  12.686250  & $-$73.26806   &   SMC100.2.114    &  15.22  &  15.39 &  0.708424  &  W04   \\
8  & 2dFS1498                                 &  B0-5II                       &  15.177833  & $-$72.41572   &   SMC108.1.21491  &  15.23  &  15.45 &  0.797337  &  I04     \\
9  & 2dFS3448                                 &  B0-5II                       &  20.075375  & $-$72.31481   &   SMC123.1.10     &  14.89  &  14.96 &  1.082169  &  \\
10 & 2dFS3395                                 &  B0-5IIIe                     &  19.869625  & $-$73.41081   &   SMC121.7.101    &  15.90  &  16.02 &  0.363367  &       \\
\hline
\end{tabular} 

\tablefoot{Table~\ref{tab06} is available in its entirety in the electronic version of the Journal. A portion is shown here for guidance regarding its form \\ and content. \\ \textbf{References:} (W04) \citet{Wyrz04}; (D08) \citet{Diago08}; (I04) \citet{Ita04}}

} 

} 
\end{table*} 


\begin{table*}
{\tiny
\caption[]{Long-period stars in the SMC}
\label{tab07}
\begin{tabular}{lllcclrrrrc}
\hline \hline
\# & Star name & Spectral type & R.A. (J2000)  & Dec. (J2000) & OGLE-III ID & $<$I$>$ & $<$V$>$ & Period (days) & Note\\ 
\hline
34 & AzV141;2dFS0915                           &  B5II                  &   13.289958  & $-$72.56311  &    SMC101.2.48081 &   14.40 &   14.47   &    31.773741  &   I04    \\
35 & 2dFS0096                                  &  B5III                 &    8.693208  & $-$73.39900  &    SMC130.2.209   &   16.91 &   17.03   &   174.907623  &  DPV(+5.1775914d) \\
36 & 2dFS0106                                  &  B8II                  &    8.762917  & $-$73.61633  &    SMC130.1.87    &   15.28 &   15.43   &    26.101492  &   \\
37 & 2dFS0344                                  &  B9II                  &   10.234250  & $-$73.12197  &    SMC125.5.9333  &   16.64 &   16.68   &     3.555301  &   \\
38 & AzV241;2dFS1437                           &  B9Ibe                 &   15.003292  & $-$72.92311  &    SMC105.1.28748 &   14.18 &   14.41   &    35.308189  &   I04, M02    \\
39 & 2dFS0573                                  &  A0II                  &   11.427792  & $-$73.57572  &    SMC125.1.14657 &   16.24 &   16.29   &    13.125437  &    \\
40 & 2dFS3113                                  &  A0II                  &   18.909042  & $-$72.49914  &    SMC115.4.5862  &   16.08 &   16.21   &    15.501976  &   \\
41 & 2dFS2073                                  &  A3Iae                 &   16.468500  & $-$72.59606  &    SMC110.3.12876 &   16.03 &   16.64   &    28.176533  &   \\
42 & SMC3-12                                   &  A3III                 &   13.496750  & $-$72.58581  &    SMC101.2.48095 &   14.05 &   14.46   &    72.918435  &  I04, M02 \\ 
43 & 2dFS2354                                  &  A5II                  &   16.991542  & $-$72.54867  &    SMC110.4.6944  &   14.65 &   14.96   &    33.677862  &   \\
\hline
\end{tabular} 

\tablefoot{Table~\ref{tab07} is available in its entirety in the electronic version of the Journal. A portion is shown here for guidance regarding its form \\ and content. \\ \textbf{References:} (I04) \citet{Ita04}; (M02) \citet{Menn02} \\ \textbf{Key:} (DPV) Double-periodic variable}
}
\end{table*} 


\begin{table*}
{\tiny
\caption[]{Long-period stars with extra variability in the SMC}
\label{tab08}
\begin{tabular}{lllcclrrrc}
\hline \hline
\# & Star name & Spectral type & R.A. (J2000)  & Dec. (J2000) & OGLE-III ID & $<$I$>$ & $<$V$>$ & Period (days) & Note\\ 
\hline
1  & 2dFS5106                &  O6-9              &    15.578500  &  $-$72.29450  &   SMC113.7.6154   &  15.19  &  15.33 &   56.007555  &      \\
2  & 2dFS1772                &  O9III-V           &    15.874167  &  $-$72.29850  &   SMC113.7.11951  &  15.05  &  15.07 &    7.009737  &      \\
3  & J005252.1-721715        &  Be XRB            &    13.217167  &  $-$72.28769  &   SMC101.4.25552  &  16.75  &  16.69        &   45.908521  & \\
4  & 2dFS2064                &  B0IV              &    16.443292  &  $-$72.11403  &   SMC113.3.21     &  14.77  &  14.95        &   27.821916  & \\
5  & 2dFS2248                &  B0-5V             &    16.775125  &  $-$73.07992  &   SMC111.4.9965   &  16.83  &  16.86 &    4.420608  &      \\
6  & 2dFS1558                &  B0-5Ve            &    15.329042  &  $-$73.85139  &   SMC107.3.9244   &  17.54  &  17.52        &    3.433239  &      \\
7  & SMC3-16                 &  B0IIIe            &    14.552417  &  $-$72.51347  &   SMC105.4.19191  &  14.55  &  14.85 &   59.631012  &  I04, M02  \\
8  & NGC330-070;SMC5\_077231 &  B0.5e             &    14.259125  &  $-$72.43204  &   SMC108.8.26225  &  15.96  &  15.99        &  586.438867  &  I04, M02  \\
9  & 2dFS5020                &  B1-3II            &    12.127417  &  $-$72.92975  &   SMC100.5.14632  &  14.33  &  14.35 &   13.706056  &      \\
10 & 2dFS0690;J0048.5-7302   &  B1.5e XRB         &    12.142750  &  $-$73.04228  &   SMC100.6.45825  &  14.73  &  14.89 &  414.913462  &      \\
\hline
\end{tabular} 

\tablefoot{Table~\ref{tab08} is available in its entirety in the electronic version of the Journal. 
A portion is shown here for guidance regarding its form \\ and content.\\ \textbf{References:} (I04) \citet{Ita04}; (M02) \citet{Menn02}}           
}
\end{table*} 


\begin{sidewaystable*}
{\tiny
\caption[]{Irregular variables in the SMC}
\label{tab09}
{\centering
\begin{tabular}{lllcclrrrc}
\hline \hline
\# & Star name & Spectral type & R.A. (J2000)  & Dec. (J2000) & OGLE-III ID & $<$I$>$ & $<$V$>$ & $\sigma_I$ & Note\\ 
\hline
182& 2dFS0839;0050-727(SMCX-3)                                  & B1-1.5IV-Ve XRB           & 13.024042 & $-$72.43422 &  SMC101.3.34265 &  14.79 &  14.92 & 0.039  &      W06                     \\
183& J0047.3-7312                                               & B1-1.5Ve XRB              & 11.848750 & $-$73.20747 &  SMC100.7.42573 &  15.87 &  16.05 & 0.044  &                                \\
184& NGC346-110;SMC5\_005120                                    & B1-2(Be-Fe)               & 14.547875 & $-$72.24519 &  SMC108.2.131   &  16.10 &  16.09 & 0.021  &      SB*, monotonic, K99             \\
185& NMC37;MPG482;KWB346-85;flames1013;SMC5\_030226             & B1-2(Be-Fe)               & 14.776404 & $-$72.16557 &  SMC108.2.37483 &  14.28 &  14.29 & 0.067  &      fading, K99                     \\
186& NGC346-073;SMC5\_025100                                    & B1-2(Be-Fe)               & 15.021833 & $-$72.26543 &  SMC108.2.16169 &  15.71 &  15.71 & 0.015  &                              \\
187& J0057.8-7207;NGC346-067;SMC5\_001335                       & B1-2(Be-Fe) XRB           & 14.460000 & $-$72.13222 &  SMC108.3.32    &  15.42 &  15.68 & 0.023  &                              \\
188& 2dFS0424                                                   & B1-2II                    & 10.633958 & $-$73.36728 &  SMC125.2.28056 &  15.25 &  15.30 & 0.267  &       bumper                  \\
189& 2dFS0876                                                   & B1-2II                    & 13.145792 & $-$72.76358 &  SMC101.1.46796 &  14.50 &  14.74 & 0.343  &       monotonic                       \\
190& 2dFS5107                                                   & B1-2II                    & 15.761333 & $-$72.08103 &  SMC113.6.27864 &  14.11 &  13.96 & 0.015  &       flicker, M02                    \\
191& AzV397;2dFS2113                                            & B1-2II                    & 16.526333 & $-$72.30097 &  SMC113.2.4432  &  13.49 &  13.64 & 0.044  &                               \\
\hline
\end{tabular}

\tablefoot{Table~\ref{tab09} is available in its entirety in the electronic version of the Journal. A portion is shown here for guidance
regarding its form and content. \\ \textbf{References:} (M02) \citet{Menn02}; (K99) \citet{Keller99}; (S09) \citet{Samus09}; (I04) \citet{Ita04}; (W06) \citet{Wat06}; (S11) \citet{Sosz11} \\ \textbf{Keys:} (SAT) Possible Saturation Noise, (FOR) Possible Foreground Star, (*)~\cite{Evans06}}
 
} 

} 

\end{sidewaystable*} 


\begin{table*}
\centering
\caption[]{Lightcurves of variables}
\label{tab10}
\begin{tabular}{lcccc}
\hline \hline
OGLE-III ID & Filter & HJD$-$2450000 (days) & Magnitude (mag) & Uncertainty (mag) \\
\hline
SMC100.1.15162 &I& 2085.90914 &15.081 &0.006  \\
SMC100.1.15162 &I& 2086.88843 &15.076 &0.005  \\
SMC100.1.15162 &I& 2103.89525 &15.219 &0.005  \\
SMC100.1.15162 &I& 2105.92315 &15.052 &0.005  \\
SMC100.1.15162 &I& 2106.89098 &15.118 &0.005  \\
SMC100.1.15162 &I& 2112.84336 &15.212 &0.005  \\
SMC100.1.15162 &I& 2116.87111 &15.206 &0.005  \\
SMC100.1.15162 &I& 2123.81803 &15.234 &0.005  \\
SMC100.1.15162 &I& 2128.82553 &15.171 &0.005  \\
SMC100.1.15162 &I& 2129.85750 &15.048 &0.005  \\
\hline
\end{tabular}

\tablefoot{Table~\ref{tab10} is available in its entirety at the CDS. A portion is shown here for guidance
regarding its form and content.}

\end{table*} 


\begin{table*}
\centering
\caption[]{Distribution of variability} 
\label{tab11}
\begin{tabular}{lcccc|cc}
\hline \hline
Type of variability & O-type & Early B & Late B & AFG-type & Early Be & Late Be \\ 
\hline
Constant & $44.3\%$ & $55.1\%$ & $65.3\%$ & $71.6\%$ & $18\%$ & $33.8\%$ \\
Low-amplitude & $28.9\%$ & $16\%$ & $23\%$ & $15.2\%$ & $14.8\%$ & $24.2\%$ \\
Irregular & $11.4\%$ & $14.1\%$ & $3.8\%$ & $0.5\%$ & $45.2\%$ & $22.6\%$ \\
Short-periodic & $4.1\%$ & $6.1\%$ & $2.1\%$ & -- & $18\%$ & $11.3\%$ \\
Long-periodic & $1.6\%$ & $1.4\%$ & $1.7\%$ & $0.8\%$ & $2.7\%$ & $3.2\%$ \\
Rotating cand. & $0.4\%$ & $1.4\%$ & $1.5\%$ & $0.2\%$ & -- & $3.2\%$ \\
Eclipsing binaries & $9.3\%$ & $5.9\%$ & $2.6\%$ & $0.4\%$ & $1.1\%$ & $1.6\%$ \\
Cepheids & -- & -- & -- & $11.4\%$ & -- & -- \\
\hline
\end{tabular}

\tablefoot{The first four columns correspond to the fractions of variability type presented in the pie charts of Fig.~\ref{fig12}. The fractions in the right two columns correspond to the pie charts of Fig.~\ref{fig14}.}   
\end{table*} 

\clearpage


\clearpage
\Online

\begin{appendix}

\onllongtab{

{\tiny
\label{tab01f}
\begin{landscape}
\begin{longtable}{lllccrrlrrrc}
\caption[]{New eclipsing binaries in the SMC}\\
\hline\hline
\# & Star Name & Spectral Type & R.A. (J2000)   & Dec. (J2000)    & Skewness & Kurtosis & OGLE-III ID & $<$I$>$ & $<$V$>$ & Period (days) & Note \\
\hline
\endfirsthead
\caption{continued.}\\
\hline
\# & Star Name & Spectral Type & R.A. (J2000)   & Dec. (J2000)    & Skewness & Kurtosis & OGLE-III ID & $<$I$>$ & $<$V$>$ & Period (days) & Note \\
\hline
\endhead
\hline
\endfoot
1   & AzV476                                   & O6.5V                      & 18.426875 & $-$73.29153 &  0.775 &  2.030  &   SMC116.6.11033 & 13.57 & 13.54 &   9.366390 &       \\
2   & WB4;NMC30;MPG342;flames1010;SMC5\_082080 & O6:V-III((f))              & 14.750221 & $-$72.17718 &  1.842 &  2.212  &   SMC108.2.37477 & 13.86 & 13.63 &   2.354849 &  SB2* \\ 
3   & 2dFS2253                                 & O7V                        & 16.780917 & $-$73.18614 &  0.528 &$-$0.989 &   SMC111.3.9765  & 15.26 & 15.09 &   1.203855 &  	 \\
4   & AzV429                                   & O7V                        & 16.999375 & $-$72.01500 &  0.365 &$-$0.257 &   SMC113.3.24653 & 14.85 & 14.68 &   2.190387   &	 \\
5   & 2dFS0404                                 & O8.5V                      & 10.541417 & $-$73.23219 &  0.760 &  2.289  &   SMC125.3.34225 & 14.95 & 14.72 &  27.281776 &	 \\
6   & NMC19;MPG445;flames1060;SMC5\_029297     & O8.5V                      & 14.769942 & $-$72.18413 &  4.800 & 31.018  &   SMC108.2.37529 & 15.40 & 15.31 &  11.833286 &  SB1* \\ 
7   & 2dFS5006                                 & O9.5III                    & 11.583333 & $-$73.10869 &  0.764 &  3.788  &   SMC100.6.61    & 15.60 & 15.44 &  32.889060 &       \\
8   & 2dFS3700                                 & O9V                        & 21.178125 & $-$73.15092 &  1.604 &  1.299  &   SMC121.4.1716  & 15.44 & 15.22 &   1.470451 &	 \\
\end{longtable}

\textbf{Keys:} (AM) Apsidal Motion, (TEB) Transient EB, (DPV) Double-Periodic Variable, (*)~Hunter et al. (in preparation), 
(**)~\cite{Evans06} 
\\$^a$Period corresponds to HJD range 2452000$-$2454200

\end{landscape}

}

} 

\clearpage
\onllongtab{

{\tiny
\label{tab02f}
\begin{landscape}
\begin{longtable}{lllccrrlrrrrc}
\caption[]{Known eclipsing binaries in the SMC}\\
\hline\hline
\# & Star Name & Spectral Type & R.A. (J2000)   & Dec. (J2000)    & Skewness & Kurtosis & OGLE-III ID & $<$I$>$ & $<$V$>$ & Reference & Period (days) & Note \\
\hline
\endfirsthead
\caption{continued.}\\
\hline
\# & Star Name & Spectral Type & R.A. (J2000)   & Dec. (J2000)    & Skewness & Kurtosis & OGLE-III ID & $<$I$>$ & $<$V$>$ & Reference & Period (days) & Note \\
\hline
\endhead
\hline
\endfoot
1   & SMC-WR7;AzV336a                      & WN4+O6I(f)        & 15.899708 & $-$72.05611 & 0.103 & $-$0.725 &  SMC113.6.33347 & 13.34 & 13.21 & N02 &  19.562969 &   SB2  \\
2   & OGLE-td9175323                       & O6.5+O5III(f)     & 15.838625 & $-$72.09383 & 0.340 & $-$1.248 &  SMC113.6.11293 & 13.82 & 13.60 & H03 &   2.206106 &     \\ 
3   & 2dFS0610                             & O7-8V             & 11.667458 & $-$73.52136 & 1.747 &    2.671 &  SMC103.5.58    & 15.66 & 15.52 & W04 &   3.320987 &     \\  
4   & OGLE-PSFid8209964                    & O8+O9             & 15.066875 & $-$72.21219 & 0.606 & $-$0.584 &  SMC108.2.46275 & 14.54 & 14.40 & H05 &   3.314737 &     \\ 
5   & OGLE-PSFid5243188                    & O8.5+O9-B0        & 12.828250 & $-$73.50439 & 1.211 &    0.276 &  SMC103.4.7254  & 15.13 & 14.95 & H05 &   1.871740 &     \\ 
6   & 2dFS1749                             & O8.5V             & 15.808500 & $-$72.41864 & 1.345 &    1.220 &  SMC113.8.11741 & 14.74 & 14.55 & W04 &   1.166967 &     \\
7   & OGLE-PSFid9010098                    & O9+B0             & 15.220417 & $-$72.79683 & 0.690 & $-$0.871 &  SMC105.2.32093 & 15.67 & 15.45 & H05 &   1.112076 &     \\ 
8   & OGLE-PSFid5095194                    & O9+O9             & 12.460083 & $-$73.32544 & 1.252 &    1.343 &  SMC100.8.52871 & 14.63 & 14.55 & H05 &   3.187497 &     \\
\end{longtable}
       
\textbf{References:} (W04) \citet{Wyrz04}, (F07) \citet{Faccioli07}, (N02) \citet{Niemela02}, (U98) \citet{Udalski98}, 
(G05) \citet{Groen05}\\ (H03) \citet{Harries03}, (H05) \citet{Hild05}, (B02) \citet{Bayne02}, (S09) \citet{Samus09},(M06) \citet{Menn06} \\ \textbf{Keys:} (AM) Apsidal Motion, (TEB) Transient EB, (*)~Hunter et al. (in preparation), (**)~\cite{Evans06} \\
$^a$Additional transient EB with $P=5.242314$ days, in the HJD range 2453200$-$2453800

\end{landscape}
}

}

\clearpage
\onllongtab{

{\tiny
\label{tab03f}
\begin{landscape}
\begin{longtable}{lllcclrrrc}
\caption[]{Rotating variables / Overcontact binaries} \\
\hline\hline
\# & Star Name & Spectral Type & R.A. (J2000)   & Dec. (J2000) & OGLE-III ID & $<$I$>$ & $<$V$>$ & Period (days) & Note \\
\hline
\endfirsthead
\caption{continued.}\\
\hline
\# & Star Name & Spectral Type & R.A. (J2000)   & Dec. (J2000) & OGLE-III ID & $<$I$>$ & $<$V$>$ & Period (days) & Note \\
\hline
\endhead
\hline
\endfoot
1   & AzV387;2dFS2050                            &  O9.5III                  &  16.416458 &   $-$72.26822 &     SMC113.2.8     &   14.11  &  13.87 &     2.972105  & 			     \\ 
2   & 2dFS2698                                   &  B0-5II                   &  17.676917 &   $-$73.32697 &     SMC116.6.129   &   15.85  &  15.71 &     6.405136  &      \\
3   & 2dFS0300                                   &  B0-5IV                   &  10.037958 &   $-$73.64939 &     SMC128.5.15134 &   16.63  &  16.44 &     2.368890  &   W04		     \\ 
4   & 2dFS2455                                   &  B0-5IV                   &  17.176375 &   $-$72.32797 &     SMC113.1.31538 &   16.02  &  15.94 &     3.528916  &   W04		     \\ 
5   & 2dFS0359                                   &  B0-5IV                   &  10.308958 &   $-$73.42589 &     SMC125.7.14593 &   16.55  &  16.69 &     1.667506  &                      \\
6   & 2dFS0010                                   &  B0-5V                    &   7.349375 &   $-$73.73997 &     SMC138.4.5904  &   17.32  &  17.30 &     3.561880  & 			     \\ 
7   & 2dFS1421                                   &  B0-5V                    &  14.956333 &   $-$72.54781 &     SMC105.4.5344  &   16.49  &  16.40 &     1.320202  & 			     \\ 
8   & 2dFS2295                                   &  B0-5V                    &  16.875167 &   $-$73.43469 &     SMC111.2.2145  &   16.56  &  16.51 &     4.326558  & 			     \\
\end{longtable} 

\textbf{References:} (W04) \citet{Wyrz04}; (D08) \citet{Diago08}; (U98) \citet{Udalski98}; (M10) \citet{Menn10} \\
\textbf{Keys:}  (*)~Hunter et al. (in prep.), (**)~\cite{Evans06}, (***)~\cite{Martayan07} 

\end{landscape} 

}

}

\clearpage
\onllongtab{

{\tiny
\label{tab04f}
\begin{landscape}
\begin{longtable}{lllccrrlrrr}
\caption[]{Known Cepheids in the SMC}\\
\hline\hline
\# & Star Name & Spectral Type & R.A. (J2000)   & Dec. (J2000) & Skewness & Kurtosis & OGLE-III ID & $<$I$>$ & $<$V$>$ & Period (days) \\
\hline
\endfirsthead
\caption{continued.}\\
\hline
\# & Star Name & Spectral Type & R.A. (J2000)   & Dec. (J2000) & Skewness & Kurtosis & OGLE-III ID & $<$I$>$ & $<$V$>$ & Period (days) \\
\hline
\endhead
\hline
\endfoot
1  & 2dFS1914           & A0II                 &  16.165833 &$-$72.31958 &$-$0.227 &$-$1.049 & SMC113.8.39669 &  15.45 & 16.12  &  3.35206   \\
2  & 2dFS1472           & A2Iab	               &  15.115958 &$-$73.37806 &$-$0.271 &$-$0.976 & SMC106.2.30024 &  16.38 & 17.10  &  2.18004   \\
3  & 2dFS1719           & A2Ib                 &  15.737167 &$-$73.41747 &  0.087  &$-$1.341 & SMC111.7.58    &  15.51 & 16.26  &  5.19790   \\
4  & 2dFS1269           & A3Ib                 &  14.467083 &$-$74.10453 &$-$0.103 &$-$1.429 & SMC107.8.8007  &  16.75 & 17.17  &  1.08680   \\
5  & 2dFS2327           & A3Ib                 &  16.936458 &$-$73.19008 &$-$0.048 &$-$1.400 & SMC111.3.12630 &  16.22 & 16.74  &  1.54276   \\
6  & 2dFS2460           & A3Ib                 &  17.185833 &$-$72.69247 &$-$0.316 &$-$1.027 & SMC110.3.10311 &  15.63 & 16.45  &  3.71093   \\
7  & 2dFS0667           & A3II                 &  11.973833 &$-$72.83844 &  0.124  &$-$1.033 & SMC100.5.37708 &  15.98 & 16.41  &  3.28218   \\
8  & 2dFS0960           & A3II                 &  13.460917 &$-$74.02661 &$-$0.526 &$-$0.817 & SMC103.1.7083  &  16.11 & 16.80  &  2.70204   \\
\end{longtable} 
      
\end{landscape} 

}
 
}

\clearpage
\onllongtab{

{\tiny
\label{tab05f}
\begin{landscape}
\begin{longtable}{lllcclrrrc}
\caption[]{Short period stars in the SMC}\\
\hline\hline
\# & Star Name & Spectral Type & R.A. (J2000)  & Dec. (J2000) & OGLE-III ID & $<$I$>$ & $<$V$>$ & Period (days) & Note \\
\hline
\endfirsthead
\caption{continued.}\\
\hline
\# & Star Name & Spectral Type & R.A. (J2000)  & Dec. (J2000) & OGLE-III ID & $<$I$>$ & $<$V$>$ & Period (days) & Note \\
\hline
\endhead
\hline
\endfoot
1  & AzV461;2dFS2772                                    &  O8V                          &   17.856583 & $-$72.16356   &   SMC118.7.8480   & 14.89  &  14.65   &    0.902841 &   	      \\
2  & 2dFS1412                                           &  O9.5III                      &   14.933875 & $-$72.75817   &   SMC105.2.26338  & 15.04  &  14.88   &    0.786212 & 	      \\
3  & 2dFS5047                                           &  O9.5V                        &   12.795125 & $-$72.80150   &   SMC101.1.7787   & 14.98  &  14.83   &    1.404488 & 	      \\
4  & MPG468                                             &  O9V                          &   14.774917 & $-$72.17486   &   SMC108.2.37536  & 15.66  &  15.46   &    2.691085 & 	      \\
5  & NGC330-077;SMC5\_000166                          	&  B0-3(Be-Fe)                  &   13.930083 & $-$72.51361   &   SMC105.5.31280  & 15.86  &  16.08   &    0.258292 &     W04   \\
6  & 2dFS5070                                           &  B0-3III                      &   13.406083 & $-$72.50981   &   SMC101.3.18041  & 15.31  &  15.15   &    0.687057 & 	      \\
7  & 2dFS0022                                           &  B0-5II                       &    7.749292 & $-$73.25556   &   SMC130.6.29     & 14.73  &  14.80   &    0.853166 & 	      \\
8  & 2dFS2247                                           &  B0-5II                       &   16.774750 & $-$72.71592   &   SMC110.3.3462   & 14.30  &  14.26   &    2.239230 & 	      \\
\end{longtable} 

\textbf{References:}  (W04) \citet{Wyrz04}; (D08) \citet{Diago08}; (I04) \citet{Ita04}  \\
\textbf{Key:} (BCEP) Candidate $\beta$ Cephei        

\end{landscape}

}

}

\clearpage
\onllongtab{

{\tiny
\label{tab06f}
\begin{landscape}
\begin{longtable}{lllcclrrrc}
\caption[]{Short period stars with extra variability in the SMC}\\
\hline\hline
\# & Star Name & Spectral Type & R.A. (J2000)  & Dec. (J2000) & OGLE-III ID & $<$I$>$ & $<$V$>$ & Period (days) & Note \\
\hline
\endfirsthead
\caption{continued.}\\
\hline
\# & Star Name & Spectral Type & R.A. (J2000)  & Dec. (J2000) & OGLE-III ID & $<$I$>$ & $<$V$>$ & Period (days) & Note \\
\hline
\endhead
\hline
\endfoot
1  & AzV480;2dFS3047                          &  O4-7Ve                       &  18.729292  & $-$72.36058   &   SMC118.1.5302   &  14.24  &  14.41 &  0.637494  & 	\\
2  & 2dFS2553                                 &  O6.5IIf                      &  17.341583  & $-$73.26150   &   SMC111.3.7734   &  15.01  &  15.02 &  0.823163  & 	\\
3  & 2dFS3357                                 &  O9.5III-V                    &  19.725167  & $-$73.16056   &   SMC121.5.11     &  14.27  &  14.38 &  0.812373  & 	\\
4  & J005517.9-723853                         &  O9.5Ve XRB                   &  13.824583  & $-$72.64806   &   SMC105.6.33072  &  15.71  &  16.02 &  0.650438  & 	\\
5  & AzV322                                   &  O9II                         &  15.750042  & $-$72.42761   &   SMC113.8.6165   &  13.71  &  13.80 &  2.586981  & 	\\
6  & AzV162                                   &  O9V                          &  13.753292  & $-$72.93822   &   SMC105.8.34986  &  13.53  &  13.75 &  1.598756  &  W04	\\
7  & J0050.7-7316                             &  B0-0.5Ve XRB                 &  12.686250  & $-$73.26806   &   SMC100.2.114    &  15.22  &  15.39 &  0.708424  &  W04   \\
8  & 2dFS1498                                 &  B0-5II                       &  15.177833  & $-$72.41572   &   SMC108.1.21491  &  15.23  &  15.45 &  0.797337  &  I04	\\
\end{longtable} 

\textbf{References:} (W04) \citet{Wyrz04}; (D08) \citet{Diago08}; (I04) \citet{Ita04}          
\end{landscape}

}
}

\clearpage
\onllongtab{

{\tiny
\label{tab07f}
\begin{landscape}
\begin{longtable}{lllcclrrrc}
\caption[]{Long period stars in the SMC}\\
\hline\hline
\# & Star Name & Spectral Type & R.A. (J2000)  & Dec. (J2000) & OGLE-III ID & $<$I$>$ & $<$V$>$ & Period (days) & Note \\
\hline
\endfirsthead
\caption{continued.}\\
\hline
\# & Star Name & Spectral Type & R.A. (J2000)  & Dec. (J2000) & OGLE-III ID & $<$I$>$ & $<$V$>$ & Period (days) & Note \\
\hline
\endhead
\hline
\endfoot
1  & Hodge53h53-137                            &  O8.5III               &   13.189333  & $-$72.65331  &    SMC101.2.14390 &   15.48 &   15.31   &     7.817038  & 	    \\
2  & AzV402;2dFS2139                           &  O9.7Iab               &   16.571708  & $-$72.46297  &    SMC110.4.12714 &   13.39 &   13.35   &   193.738651  & 	 \\
3  & 2dFS0658                                  &  B0-5II              	&   11.899208  & $-$72.96194  &    SMC100.5.6784  &   14.21 &   14.13   &   102.880537  & 	 \\
4  & 2dFS3211                                  &  B0-5II              	&   19.179083  & $-$73.35328  &    SMC116.2.6414  &   15.41 &   15.21   &    20.446851  & 	 \\
5  & 2dFS2032                                  &  B0-5III             	&   16.383125  & $-$72.18322  &    SMC113.2.16438 &   15.94 &   15.80   &    54.525955  & 	 \\
6  & 2dFS1126                                  &  B0-5IV              	&   14.029375  & $-$73.00061  &    SMC105.8.15194 &   15.91 &   16.02   &   187.871723  &   I04, M02, DPV(+3.111001d)     \\
7  & 2dFS1148                                  &  B0-5IV              	&   14.124292  & $-$73.02597  &    SMC105.8.15213 &   16.16 &   16.09   &    55.153698  &         \\
8  & 2dFS0852                                  &  B0-5IV	        &   13.055042  & $-$72.38642  &    SMC101.4.8405  &   16.49 &   16.68   &    43.307767  &        \\
\end{longtable} 

\textbf{References:} (I04) \citet{Ita04},  (M02) \citet{Menn02}   \\
\textbf{Key:} (DPV) Double-Periodic Variable
\end{landscape}

}
}

\clearpage
\onllongtab{

{\tiny
\label{tab08f}
\begin{landscape}
\begin{longtable}{lllcclrrrc}
\caption[]{Long period stars with extra variability in the SMC}\\
\hline\hline
\# & Star Name & Spectral Type & R.A. (J2000)  & Dec. (J2000) & OGLE-III ID & $<$I$>$ & $<$V$>$ & Period (days) & Note \\
\hline
\endfirsthead
\caption{continued.}\\
\hline
\# & Star Name & Spectral Type & R.A. (J2000)  & Dec. (J2000) & OGLE-III ID & $<$I$>$ & $<$V$>$ & Period (days) & Note \\
\hline
\endhead
\hline
\endfoot
1  & 2dFS5106                &  O6-9              &    15.578500  &  $-$72.29450  &   SMC113.7.6154   &  15.19  &  15.33	&   56.007555  &      \\
2  & 2dFS1772                &  O9III-V           &    15.874167  &  $-$72.29850  &   SMC113.7.11951  &  15.05  &  15.07	&    7.009737  &      \\
3  & J005252.1-721715        &  Be XRB            &    13.217167  &  $-$72.28769  &   SMC101.4.25552  &  16.75  &  16.69        &   45.908521  &	\\
4  & 2dFS2064                &  B0IV              &    16.443292  &  $-$72.11403  &   SMC113.3.21     &  14.77  &  14.95        &   27.821916  &	\\
5  & 2dFS2248                &  B0-5V             &    16.775125  &  $-$73.07992  &   SMC111.4.9965   &  16.83  &  16.86	&    4.420608  &      \\
6  & 2dFS1558                &  B0-5Ve            &    15.329042  &  $-$73.85139  &   SMC107.3.9244   &  17.54  &  17.52        &    3.433239  &      \\
7  & SMC3-16                 &  B0IIIe            &    14.552417  &  $-$72.51347  &   SMC105.4.19191  &  14.55  &  14.85	&   59.631012  &  I04, M02  \\
8  & NGC330-070;SMC5\_077231 &  B0.5e             &    14.259125  &  $-$72.43204  &   SMC108.8.26225  &  15.96  &  15.99        &  586.438867  &  I04, M02  \\
\end{longtable} 

\textbf{References:} (I04) \citet{Ita04}; (M02) \citet{Menn02} \\
\end{landscape}

}

}

\clearpage
\onllongtab{

{\tiny
\label{tab09f}
\begin{landscape}
\begin{longtable}{lllcclrrrc}
\caption[]{Irregular variables in the SMC}\\
\hline\hline
\# & Star Name & Spectral Type & R.A. (J2000)  & Dec. (J2000) & OGLE-III ID & $<$I$>$ & $<$V$>$ & $\sigma_I$ & Note \\
\hline
\endfirsthead
\caption{continued.}\\
\hline
\# & Star Name & Spectral Type & R.A. (J2000)  & Dec. (J2000) & OGLE-III ID & $<$I$>$ & $<$V$>$ & $\sigma_I$ & Note \\
\hline
\endhead
\hline
\endfoot
1  & SMC-WR1;AzV2a                                          	& WN3ha                     & 10.926292 & $-$73.48203 &  SMC125.1.25666 &  15.23 &  15.27 & 0.016  &	S09		       \\
2  & SMC-WR4;AzV81;Sk41                                     	& WN6h                      & 12.680958 & $-$73.45158 &  SMC103.4.25983 &  13.62 &  13.70 & 0.027  &       M02                       \\
3  & AzV75                                                      & O5.5I(f)                  & 12.635000 & $-$72.87678 &  SMC100.4.37674 &  12.84 &  99.99 & 0.015  &	SAT		       \\
4  & AzV61;2dFS0748     	 	                        & O5.5III((f))              & 12.507750 & $-$72.19072 &  SMC102.8.8466  &  13.82 &  13.63 & 0.039  &     	S09                     \\
5  & 2dFS2898           	 	                        & O6-9III-V                 & 18.224625 & $-$73.50719 &  SMC116.8.7811  &  14.51 &  14.73 & 0.016  &       			\\
6  & 2dFS0936           	 	                        & O6.5f?p                   & 13.375250 & $-$72.69564 &  SMC101.2.21946 &  14.28 &  14.09 & 0.015  & 				\\
7  & 2dFS5001           	 	                        & O6V                       & 11.238000 & $-$73.01294 &  SMC126.1.18639 &  14.48 &  14.58 & 0.115  & 	M02			\\
8  & 2dFS3157           	 	                        & O7-9V                     & 19.002167 & $-$72.72397 &  SMC115.3.6580  &  15.19 &  15.44 & 0.025  & 				\\
\end{longtable} 

\textbf{References:} (M02) \citet{Menn02}; (K99) \citet{Keller99}; (S09) \citet{Samus09}; (I04) \citet{Ita04}; (W06) \citet{Wat06}; (S11) \citet{Sosz11}\\
\textbf{Keys:} (SAT) Possible Saturation Noise, (FOR) Possible Foreground Star, (*)~\cite{Evans06}

\end{landscape}

}

}

\end{appendix}

\end{document}